\documentclass[twocolumn]{aastex631}
\pdfoutput=1

\defcitealias{Nutter2007}{NWT07}

\newcommand{\tnm}[1]{\tablenotemark{#1}}

\begin{document}

\title{An ALMA search for substructure and fragmentation in starless cores in Orion B North}

\author[0009-0001-8625-505X]{Samuel D. Fielder}
\affiliation{Department of Physics and Astronomy, University of Victoria, Victoria, BC, V8P 1A1, Canada}

\author[0000-0002-5779-8549]{Helen Kirk}
\affiliation{Herzberg Astronomy and Astrophysics Research Centre, National Research Council of Canada, 5071 West Saanich Road, Victoria, BC, V9E 2E7, Canada}
\affiliation{Department of Physics and Astronomy, University of Victoria, Victoria, BC, V8P 1A1, Canada}

\author[0000-0003-0749-9505]{Michael M. Dunham}
\affiliation{Department of Physics, State University of New York at Fredonia, Fredonia, NY 14063, USA}

\author[0000-0003-1252-9916]{Stella S. R. Offner}
\affiliation{Department of Astronomy, The University of Texas at Austin, Austin, TX 78712, USA}

\begin{abstract}
We present Atacama Large Millimeter/submillimeter Array (ALMA) Cycle 3 observations of 73 starless and protostellar cores in the Orion B North molecular cloud. We detect a total of 34 continuum sources at 106~GHz, and after comparisons with other data, 4 of these sources appear to be starless. Three of the four sources are located near groupings of protostellar sources, while one source is an isolated detection. We use synthetic observations of a simulation modeling a collapsing turbulent, magnetized core to compute the expected number of starless cores that should be detectable with our ALMA observations and find at least two (1.52) starless core should be detectable, consistent with our data. We run a simple virial analysis of the cores to put the Orion B North observations into context with similar previous ALMA surveys of cores in Chamaeleon I and Ophiuchus. We conclude that the Chamaeleon I starless core population is characteristically less bounded than the other two populations, along with external pressure contributions dominating the binding energy of the cores. These differences may explain why the Chamaeleon I cores do not follow turbulent model predictions, while the Ophiuchus and Orion B North cores are consistent with the model. 
\end{abstract}


\section{Introduction} 
\label{sec:intro}

Dense molecular cloud cores, sub-parsec scale ($<$0.2~pc) over-densities within molecular clouds, are the immediate progenitors of stars \citep{Bergin2007, DiFrancesco2007}. Star formation is an inherently multi-scale process, and we access the wide variety of scales, from cloud to core, by way of different observations. The types of structures present at each scale may provide evidence as to which physical processes are dominant and how they govern star formation. Turbulence is thought to offer global support to the overall molecular cloud collapse, while also driving local collapse at the core scale \citep{MacLow2004, Ballesteros-Paredes2007}. Studies of the role of turbulence in the transition from dense cores to individual protostars can provide a deeper understanding for hallmark results in the theory of star formation, such as the initial mass function \citep{Goodwin2008, Holman2013, Offner2014}, and protostellar multiplicity \citep{Chen2013, Lomax2015, Offner2023}.

Around half of all stars exist in binary or multiple systems \citep{Moe2017}, likely driven by some combination of disk fragmentation, core fragmentation and dynamical capture \citep[see recent review by][]{Offner2023}. Recent observational studies in the Perseus cloud \citep{Tobin2016} and the Orion cloud \citep{Tobin2020} show that there is a distinct bimodal distribution in the separations of multiple stars systems, with peaks at $\sim$75~au and $\sim$3000~au. \citet{Tobin2020} attribute the larger separation peak to core fragmentation processes, while the smaller peak is attributed to disk fragmentation processes. Additionally, simulations of star formation in turbulent molecular clouds show that a significant fraction of smaller separation multiples can be produced by significant orbital evolution following core fragmentation \citep{Lee2019, Kuruwita2023}.

\citet{Pokhrel2018} performed a multi-scale fragmentation study ranging from cloud ($\geq$ 10~pc) to core ($\sim$10~au), indicating that at all scales, the number of substructures within parent structures was lower than expected from thermal Jeans fragmentation. On the larger scales, observed separations of dense cores along filamentary structures often show inconsistencies with classical analytic cylindrical fragmentation models without the consideration of magnetic fields or turbulence \citep[e.g.,][]{Andre2014, Konyves2020}. Other studies on fragmentation on smaller scales, those within dense cores, similarly show fewer substructures than expected from a thermal Jeans model \citep[e.g.,][]{Das2021}. Meanwhile, studies such as \citet{Ohashi2018} and \citet{Palau2015} show that the separations between substructures within cores are consistent with thermal Jeans lengths. Further studies on fragmentation within dense cores are clearly needed to discern the importance of which physics are at play. The Atacama Large Millimetre/submillimetre Array (ALMA) provides an efficient avenue to search for rare very high-density peaks within starless cores, a natural indicator for thermal versus turbulent processes, where thermal-only processes take longer to form small high-density peaks \citep{Dunham2016}. Population studies, in general, through the total number of detections, can provide a basis to evaluate whether fragmentation within the core is more consistent with thermal or turbulent origins.

Large field-of-view (sub)millimeter observations have proven useful as a way of efficiently mapping entire star forming regions and their dense core populations. These dust continuum studies are good tracers of column density peaks, and can therefore characterize the locations, sizes and approximate masses of entire dense core populations \citep[e.g.,][]{Motte1998, Enoch2007, Ward-Thompson2007, Konyves2015}. Early (sub)millimetre studies indicated that starless cores have inner flat density substructures \citep{Ward-Thompson1994}, and that the density profiles are often approximated as smooth Bonnor-Ebert (BE) spheres, which are centrally flat, and roughly drop as $r^{-2}$ toward the edge \citep{Ebert1955, Bonnor1956}. However, these single-dish (sub)millimetre observations lack the sufficient angular resolution to probe the inner structures where fragmentation could be taking place. 

High resolution facilities, like interferometers, are needed to be able to resolve the structure present within dense cores, leading to a more complete look at the role of core fragmentation. The earliest large interferometric survey of starless dense cores was a CARMA (Combined Array for Research in Millimetre-wave Astronomy) survey of 12 dense cores in the Perseus molecular cloud \citep{Schnee2010}. Despite observing each core for 8 hours, no detectable substructure was found \citep{Schnee2010}. \citet{Offner2012} showed that CARMA most likely lacked the sensitivity needed to detect substructure, and that ALMA would have the necessary sensitivity to see core substructure resulting from turbulence-driven fragmentation. 

Interferometers by nature are only sensitive to a range of angular scales, the values of which are set by the observed wavelength and separations between antennas. Highly dense and compact features are best detected with larger antenna separations, with accompanying poorer sensitivity to larger-scale features. Conversely, more compact antenna configurations, like those observed with ALMA's Atacama Compact Array (ACA), are much more successful at detecting larger and less dense starless core fragments \citep[e.g.,][]{Dutta2020,Tokuda2020, Sahu2021, Sahu2023}. In cases that do not have additional larger separation data (e.g., ALMA 12m array data), there is insufficient resolution to determine whether or not there are any compact peaks within the fragments.

The first dense core population study with the ALMA 12m array, to hunt for these rare very high density substructures, was performed by \citet{Dunham2016} in the Chamaeleon I region, however, no substructures were detected for any of the 56 starless cores observed. This contradicts the expected number of detections predicted by synthetic observations generated from simulations turbulent prestellar cores. \citet{Dunham2016} also demonstrated that structures generated from turbulent fragmentation should be detected at a rate approximately 100 times higher than BE sphere modeled cores, for their specific ALMA observations. The factor of 100 less in detectability in BE spheres does not mean they could never be detected: \citet{Caselli2019}, for example, demonstrate that an evolved BE sphere model fit to the highly evolved starless core L1544 would have a central flat region only 1.75\arcsec ($\sim$250~au at a distance of 140~pc) with their ALMA 12m plus ACA observations, making a detection likely. The main difference between turbulently collapsing cores and non-turbulent, smoothly collapsing cores is the lifetime of the high-density peaks, with collapse occurring earlier in the case with turbulence \citep{Offner2012}.

Following the Chamaeleon I ALMA study, \citet{Kirk2017} performed a population study of the L1688 molecular cloud in Ophiuchus, where the predictions derived from synthetic observations of the same turbulent fragmentation simulations matched the observed two compact substructure detections.

To date, several other groups have carried out deeper observations, including the use of the ACA, to provide better sensitivity to larger-scale features \citep[e.g.,][]{Sato2023, Hirano2024}. \citet{Tokuda2020} performed observations with the ACA of 39 dense cores (32 starless and 7 protostellar) in the Taurus molecular cloud, achieving an angular resolution of 6.5\arcsec ($\sim$900~au). \citet{Tokuda2020} detected multiple complex substructures with a typical size scale of $\sim$1000~au across a total of 12 of the observed starless cores. Additionally, \citet{Sahu2021}, through the ALMA Survey of Orion Planck Galactic Cold Clumps (ALMASOP), have performed detailed studies on the substructure of five dense cores, probing much higher resolution scales down to 0.8\arcsec ($\sim$320~au). One core studied showed clear signatures of fragmentation, characterized by relatively high density ($2-8\times10^7~\text{cm}^{-3}$) and separations of $\sim$1200~au \citep{Sahu2021}, including one instance of a proto-quadruple system \citep{Luo2023}. Although the results of both \citet{Tokuda2020} and \citet{Sahu2021} indicate potential agreement with the turbulent fragmentation model, both studies do not explicitly test the number of starless core detections.

To complement the previous studies performed by \citet{Dunham2016} and \citet{Kirk2017}, we present results from a similarly designed survey of starless cores in the Orion B molecular cloud. In contrast to Chamaeleon I and Ophiuchus, Orion is the closest high-mass star-forming region, with the Orion B North portion of the complex situated at a distance of approximately 419~pc \citep{Zucker2019}. As noted earlier, the number of detections of high density peaks in \citet{Dunham2016} and \citet{Kirk2017} taken together were puzzling, with \citet{Dunham2016} showing a strong disagreement with the predictions from the turbulent simulations, while \citet{Kirk2017} agreed well. One possible explanation for this discrepancy is the dynamical state of the population of dense cores in each cloud; indeed, \citet{Tsitali2015} showed that most dense cores in Chamaeleon I appear unbound, indicating that the cores may be dispersing rather than evolving to form stars. Utilizing comparable datasets for all three regions, we analyze the boundedness of the dense core populations, and use these results to interpret the number of detections of high density peaks within Orion B North in the broader context.

In this paper, we present an ALMA 3~mm continuum survey of 73 dense cores in the Orion B North molecular cloud. In Section \ref{sec:observations}, we present the observations and data reduction applied, and in Section  \ref{sec:detections}, we perform a catalog search for any associated protostellar sources. In Section \ref{sec:derived-properties} we analyze the ALMA detections and estimate their radius, mass and number density. In Section \ref{sec:substructure-fragmentation}, we use numerical simulations of starless core evolution to create synthetic observations to predict the expected number of starless core detections in our dataset. In Section \ref{sec:comparisons}, we consider our results in the context of the previous ALMA population studies, and run a simple virial analysis to explore whether differences in core boundedness is a plausible explanation for the varied levels of agreement from the turbulent fragmentation model. We summarize our findings in Section \ref{sec:conclusions}.

\section{Observations} 
\label{sec:observations}

\subsection{Target Selection: SCUBA}
\label{sec:target-selection}

\begin{figure}
    \plotone{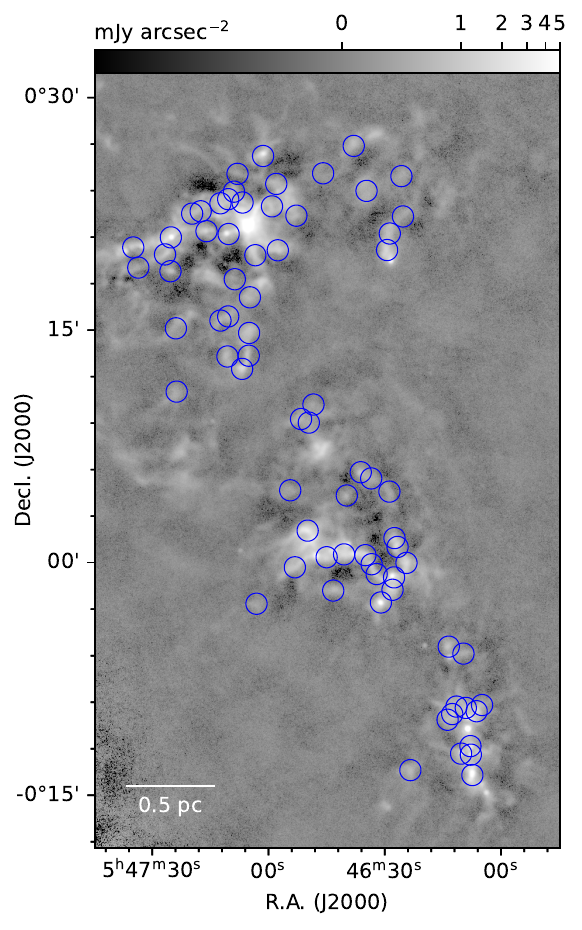}
    \caption{
    SCUBA2 850~$\mu$m image of the Orion BN cloud adapted from \citet{Kirk2016}. The blue circles show all 73 \textit{starless} dense cores identified by \citetalias{Nutter2007} which were observed by ALMA, with the diameters of the circles equal to the  primary beam of the 12~m observations.
    \label{fig:pointings}
    }
\end{figure}

In their re-analysis of all SCUBA archive data of the Orion star-forming regions, \citet{Nutter2007} (hereafter \citetalias{Nutter2007}) identified a total of 393 dense cores in the Orion molecular cloud complex. The detected cores were classified as protostellar or starless using archival data from the Spitzer Space Telescope. In the Orion B North portion of the complex, covering the area of NGC 2068/2071, 73 of the 100 dense cores were classified as starless \citepalias{Nutter2007}. The \citetalias{Nutter2007} catalog is sensitive down to a mass of $0.1M_{\odot}$ and maps the entirety of the NGC 2068/2071 star forming regions. We obtained ALMA Cycle 3 observations observations of all 73 starless cores identified by \citetalias{Nutter2007} in the Orion B North region.
Figure \ref{fig:pointings} shows the 73 ALMA pointings overlaid on a more recent SCUBA-2 850~$\mu$m image of Orion BN presented by \citet{Kirk2016}, and positional information is given in Table \ref{tab:observation-noise-levels}.

\subsection{ALMA Data}
\label{sec:alma-data}

The ALMA Band 3 data were observed between 2016 March 08 and 2016 August 27, and consisted of 7 unique observation times, using an average of 38 antennas. Four spectral windows were used, with three configured for continuum measurements, centered at 101~GHz, 103~GHz, and 113~GHz, each with a bandwidth of 1875~MHz. The total continuum bandwidth is therefore approximately 6~GHz at a central frequency of 106~GHz. The last spectral window was configured for observation of $^{12}$CO~($1-0$) emission at 115~GHz. Our main focus in this study is the continuum data; we use the CO line data appropriately when needed (see Section \ref{sec:alma-co-data} for details).

Flux and bandpass calibrators were observed at the beginning of each execution block, followed by single pointings of all 73 science targets, with periodic phase gain calibrators therein. Most sources (50) were observed during all 7 execution blocks, totalling 270~seconds of on source time, with the remaining (23) sources missing the last execution block, totalling 230~seconds of on-source time. These continuum observations were requested to have a 1$\sigma$~rms noise of $0.07~\text{mJy}~\text{beam}^{-1}$, and was achieved for all ALMA pointings (see Table \ref{tab:observation-noise-levels} and below for details).

The array was in a relatively compact configuration for the first five observation windows (C-2/3), with antenna separations ranging from approximately 15~m to 460~m. In the final two observations, the array was in a slightly more expanded configuration (C-4/5) with antenna separations ranging from approximately 15~m to 1125~m. This yields an average synthesized beam of 1.5\arcsec $\times$ 1.3\arcsec ($\sim$ 630~au $\times$ 550~au) at a position angle of 75~degrees, along with a maximum recoverable scale of approximately 14\arcsec. There is no significant difference in the synthesized beam between sources that were only observed for 230~seconds, and those observed for the full 270~seconds.

The calibration and reduction was conducted in the pipeline version 4.7.0 of the Common Astronomy Software Applications (\textsc{CASA}) software. For imaging and analysis, \textsc{CASA} version 6.5.0 was utilized for its refined automasking routines \citep{TheCASATeam2022}. 

To construct the continuum images, the line emission was first subtracted from the last spectral window in order to make use of all available continuum emission present, along with these standard choices of imaging parameters: Briggs weighting with a robust value of $R = 0.5$, a uvtaper of 0.8\arcsec to reduce sidelobe contamination, and default automasking parameters. Self-calibration routines were performed on the continuum where it improved image quality, which in our case was where the peak emission was above approximately 10~mJy~beam$^{-1}$. This brought all fields to the rms noise level requested for the observations, 0.07~mJy~beam$^{-1}$. In the case where the field of view of individual ALMA observations overlapped, mosaicking was conducted (prior to self-calibration where appropriate) to improve the final sensitivity of the image. The observations lying within a mosaicked area are noted in Table \ref{tab:observation-noise-levels}. 

For line emission imaging, the continuum was first subtracted, along with only one change with respect to the above parameters, a modified version of the Briggs weighting scheme called \texttt{briggsbwtaper}, with the same robust value of $R=0.5$. The purpose with this choice of weighting scheme is to modify the cube imaging weights to have a similar density to that of the continuum imaging weights \citep{TheCASATeam2022}, which gave the best results in our images. We use this data as an indicator for protostellar nature, as the data achieves a sensitivity of 1.1~K in a 0.5~km~s$^{-1}$ channel, which is sufficient to detect outflow signatures, if any are present near our identified detections.

All final images produced were corrected for primary beam attenuation. We subsequently use \textsc{CASA}'s \texttt{imfit} task to fit elliptical Gaussians to all continuum sources found.

\startlongtable
\begin{deluxetable*}{lcccccc}
    \tablecaption{Noise Levels of Targeted ALMA Observations \label{tab:observation-noise-levels}}
    \tablehead{
        \colhead{Field\tnm{a}} &
        \colhead{R.A.} &
        \colhead{Decl.} &
        \colhead{ALMA Mosaic Field\tnm{b}} &
        \colhead{$1\sigma$ rms\tnm{c}} &
        \colhead{Int. Time\tnm{d}} &
        \colhead{YSO\tnm{e}} \\
        \colhead{} &
        \colhead{(J2000)} &
        \colhead{(J2000)} &
        \colhead{} &
        \colhead{($\text{mJy}~\text{beam}^{-1}$)} &
        \colhead{(s)} &
        \colhead{}
    }
    \startdata
    BN-546049-00911 & 05:46:04.90 & -00:09:11.00 & BN-546049-00911Mosaic & 0.045 & 270 & N \\
    BN-546063-00935 & 05:46:06.30 & -00:09:35.00 & BN-546049-00911Mosaic & 0.045 & 270 & Y \\
    BN-546074-01342 & 05:46:07.40 & -00:13:42.00 & BN-546074-01342Mosaic & 0.061 & 270 & Y \\
    BN-546078-01223 & 05:46:07.80 & -00:12:23.00 & BN-546074-01342Mosaic & 0.061 & 270 & Y \\
    BN-546079-01150 & 05:46:07.90 & -00:11:50.00 & BN-546074-01342Mosaic & 0.061 & 270 & Y \\
    BN-546091-00922 & 05:46:09.10 & -00:09:22.00 & BN-546049-00911Mosaic & 0.045 & 270 & N \\
    BN-546097-00552 & 05:46:09.70 & -00:05:52.00 & BN-546097-00552Mosaic & 0.054 & 270 & N \\
    BN-546103-01219 & 05:46:10.30 & -00:12:19.00 & BN-546074-01342Mosaic & 0.061 & 270 & Y \\
    BN-546115-00917 & 05:46:11.50 & -00:09:17.00 & BN-546049-00911Mosaic & 0.045 & 270 & N \\
    BN-546126-00946 & 05:46:12.60 & -00:09:46.00 & BN-546049-00911Mosaic & 0.045 & 270 & N \\
    BN-546135-00525 & 05:46:13.50 & -00:05:25.00 & BN-546097-00552Mosaic & 0.054 & 270 & Y \\
    BN-546138-01008 & 05:46:13.80 & -00:10:08.00 & BN-546049-00911Mosaic & 0.045 & 270 & N \\
    BN-546234-01323 & 05:46:23.40 & -00:13:23.00 & ... & 0.050 & 270 & N \\
    BN-546244-00001 & 05:46:24.40 & -00:00:01.00 & BN-546244-00001Mosaic & 0.054 & 270 & N \\
    BN-546253+02220 & 05:46:25.30 & +00:22:20.00 & ... & 0.053 & 230 & N \\
    BN-546257+02456 & 05:46:25.70 & +00:24:56.00 & ... & 0.057 & 230 & N \\
    BN-546267+00101 & 05:46:26.70 & +00:01:01.00 & BN-546244-00001Mosaic & 0.054 & 270 & N \\
    BN-546275+00135 & 05:46:27.50 & +00:01:35.00 & BN-546244-00001Mosaic & 0.054 & 270 & N \\
    BN-546276-00057 & 05:46:27.60 & -00:00:57.00 & BN-546244-00001Mosaic &    0.076\tnm{$\ast$} &             270 &   Y \\
    BN-546280-00145 & 05:46:28.00 & -00:01:45.00 & BN-546244-00001Mosaic & 0.054 & 270 & N \\
    BN-546287+02114 & 05:46:28.70 & +00:21:14.00 & BN-546287+02114Mosaic & 0.055 & 230 & N \\
    BN-546288+00435 & 05:46:28.80 & +00:04:35.00 & ... & 0.047 & 270 & N \\
    BN-546294+02010 & 05:46:29.40 & +00:20:10.00 & BN-546287+02114Mosaic & 0.055 & 230 & N \\
    BN-546310-00234 & 05:46:31.00 & -00:02:34.00 & BN-546244-00001Mosaic & 0.054 & 270 & Y \\
    BN-546321-00044 & 05:46:32.10 & -00:00:44.00 & BN-546244-00001Mosaic & 0.054 & 270 & N \\
    BN-546334-00006 & 05:46:33.40 & -00:00:06.00 & BN-546244-00001Mosaic & 0.054 & 270 & Y \\
    BN-546335+00526 & 05:46:33.50 & +00:05:26.00 & BN-546335+00526Mosaic & 0.047 & 270 & N \\
    BN-546347+02359 & 05:46:34.70 & +00:23:59.00 & ... & 0.056 & 230 & N \\
    BN-546350+00029 & 05:46:35.00 & +00:00:29.00 & BN-546244-00001Mosaic & 0.054 & 270 & N \\
    BN-546362+00550 & 05:46:36.20 & +00:05:50.00 & BN-546335+00526Mosaic & 0.047 & 270 & N \\
    BN-546380+02653 & 05:46:38.00 & +00:26:53.00 & ... & 0.060 & 230 & N \\
    BN-546398+00420 & 05:46:39.80 & +00:04:20.00 & ... & 0.048 & 270 & Y \\
    BN-546405+00032 & 05:46:40.50 & +00:00:32.00 & BN-546405+00032Mosaic & 0.050 & 270 & N \\
    BN-546433-00148 & 05:46:43.30 & -00:01:48.00 & ... & 0.054 & 270 & N \\
    BN-546450+00021 & 05:46:45.00 & +00:00:21.00 & BN-546405+00032Mosaic & 0.050 & 270 & N \\
    BN-546459+02507 & 05:46:45.90 & +00:25:07.00 & ... & 0.059 & 230 & N \\
    BN-546484+01012 & 05:46:48.40 & +00:10:12.00 & BN-546484+01012Mosaic & 0.042 & 270 & N \\
    BN-546496+00902 & 05:46:49.60 & +00:09:02.00 & BN-546484+01012Mosaic & 0.042 & 270 & N \\
    BN-546499+00204 & 05:46:49.90 & +00:02:04.00 & ... & 0.049 & 270 & N \\
    BN-546516+00916 & 05:46:51.60 & +00:09:16.00 & BN-546484+01012Mosaic & 0.042 & 270 & N \\
    BN-546528+02223 & 05:46:52.80 & +00:22:23.00 & ... & 0.054 & 230 & N \\
    BN-546532-00018 & 05:46:53.20 & -00:00:18.00 & ... & 0.051 & 270 & N \\
    BN-546544+00440 & 05:46:54.40 & +00:04:40.00 & ... & 0.047 & 270 & N \\
    BN-546576+02009 & 05:46:57.60 & +00:20:09.00 & ... & 0.054 & 230 & N \\
    BN-546580+02426 & 05:46:58.00 & +00:24:26.00 & ... & 0.058 & 230 & N \\
    BN-546591+02259 & 05:46:59.10 & +00:22:59.00 & ... & 0.055 & 230 & N \\
    BN-547014+02614 & 05:47:01.40 & +00:26:14.00 & ... & 0.059 & 230 & Y \\
    BN-547031-00239 & 05:47:03.10 & -00:02:39.00 & ... & 0.053 & 270 & N \\
    BN-547034+01950 & 05:47:03.40 & +00:19:50.00 & ... & 0.054 & 270 & N \\
    BN-547048+01707 & 05:47:04.80 & +00:17:07.00 & ... & 0.050 & 270 & N \\
    BN-547050+01449 & 05:47:05.00 & +00:14:49.00 & ... & 0.049 & 270 & N \\
    BN-547051+01321 & 05:47:05.10 & +00:13:21.00 & BN-547051+01321Mosaic & 0.050 & 270 & N \\
    BN-547067+02314 & 05:47:06.70 & +00:23:14.00 & BN-547067+02314Mosaic & 0.049 & 230 & N \\
    BN-547068+01230 & 05:47:06.80 & +00:12:30.00 & BN-547051+01321Mosaic & 0.050 & 270 & N \\
    BN-547080+02505 & 05:47:08.00 & +00:25:05.00 & BN-547067+02314Mosaic & 0.049 & 230 & N \\
    BN-547087+01817 & 05:47:08.70 & +00:18:17.00 & ... & 0.052 & 270 & N \\
    BN-547089+02356 & 05:47:08.90 & +00:23:56.00 & BN-547067+02314Mosaic & 0.049 & 230 & N \\
    BN-547103+02112 & 05:47:10.30 & +00:21:12.00 & ... & 0.066 & 230 & Y \\
    BN-547104+01553 & 05:47:10.40 & +00:15:53.00 & BN-547104+01553Mosaic & 0.045 & 270 & N \\
    BN-547104+02327 & 05:47:10.40 & +00:23:27.00 & BN-547067+02314Mosaic & 0.049 & 230 & Y \\
    BN-547106+01318 & 05:47:10.60 & +00:13:18.00 & BN-547051+01321Mosaic & 0.050 & 270 & N \\
    BN-547124+01537 & 05:47:12.40 & +00:15:37.00 & BN-547104+01553Mosaic & 0.045 & 270 & N \\
    BN-547124+02311 & 05:47:12.40 & +00:23:11.00 & BN-547067+02314Mosaic & 0.049 & 230 & N \\
    BN-547160+02123 & 05:47:16.00 & +00:21:23.00 & BN-547160+02123Mosaic & 0.048 & 230 & Y \\
    BN-547175+02240 & 05:47:17.50 & +00:22:40.00 & BN-547160+02123Mosaic & 0.048 & 230 & N \\
    BN-547197+02231 & 05:47:19.70 & +00:22:31.00 & BN-547160+02123Mosaic & 0.048 & 230 & N \\
    BN-547237+01102 & 05:47:23.70 & +00:11:02.00 & ... & 0.048 & 270 & N \\
    BN-547239+01507 & 05:47:23.90 & +00:15:07.00 & ... & 0.051 & 270 & N \\
    BN-547252+02059 & 05:47:25.20 & +00:20:59.00 & BN-547252+02059Mosaic & 0.055 & 230 & Y \\
    BN-547253+01848 & 05:47:25.30 & +00:18:48.00 & BN-547252+02059Mosaic & 0.055 & 270 & N \\
    BN-547267+01953 & 05:47:26.70 & +00:19:53.00 & BN-547252+02059Mosaic & 0.055 & 230 & N \\
    BN-547336+01902 & 05:47:33.60 & +00:19:02.00 & BN-547336+01902Mosaic & 0.054 & 270 & N \\
    BN-547349+02020 & 05:47:34.90 & +00:20:20.00 & BN-547336+01902Mosaic & 0.054 & 230 & N \\
    \enddata
    \tablenotetext{a}{Observed SCUBA core name from \citet{Nutter2007}.}
    \tablenotetext{b}{For individual fields which overlap in coverage, the mosaic field name is given (taken to be the eastern-most field).}
    \tablenotetext{c}{1$\sigma$ root-mean-square noise, computed in non-detection areas of the (mosaicked, if applicable) field. This value was computed from the non-primary beam corrected image.}
    \tablenotetext{d}{Total integration time on the individual field.}
    \tablenotetext{e}{Protostellar classification based on more recent catalogs studied (see Section \ref{sec:associations} for details).}
    \tablenotetext{*}{Due to a bright central protostellar source, we report a more representative value of rms for this individual field, as opposed to the rms value computed for the mosaicked field.}
\end{deluxetable*}

\begin{longrotatetable}
    \begin{deluxetable*}{clrDDDDDDrDDDDrr}
    \tablecaption{Observed Properties of ALMA Detections\label{tab:observed-properties}}
    \tablehead{
        \colhead{Src} &
        \colhead{R.A.} &
        \colhead{Decl.} & 
        \multicolumn{2}{c}{Pk\tnm{a}} &
        \multicolumn{2}{c}{Pk$_{\text{err}}$\tnm{a}} &
        \multicolumn{2}{c}{Tot\tnm{a}} &
        \multicolumn{2}{c}{Tot$_{\text{err}}$\tnm{a}} &
        \multicolumn{2}{c}{FWHM$_{\text{a}}$\tnm{a}} &
        \multicolumn{2}{c}{FWHM$_{\text{b}}$\tnm{a}} &
        \colhead{P.A.\tnm{a}} &
        \multicolumn{4}{c}{FWHM$_{\text{a,d}}$\tnm{b}} &
        \multicolumn{4}{c}{FWHM$_{\text{b,d}}$\tnm{b}} &
        \multicolumn{2}{c}{P.A.$_{\text{d}}$\tnm{b}} \\
        \colhead{\#} &
        \colhead{(J2000)} &
        \colhead{(J2000)} & 
        \multicolumn{4}{c}{($\text{mJy}~\text{beam}^{-1}$)} &
        \multicolumn{4}{c}{(mJy)} &
        \multicolumn{4}{c}{(arcsec)} &
        \colhead{(deg)} &
        \multicolumn{4}{c}{(arcsec)} &
        \multicolumn{4}{c}{(arcsec)} &
        \multicolumn{2}{c}{(deg)} \\[-0.2cm]
        \colhead{} &
        \colhead{} &
        \colhead{} & 
        \multicolumn{4}{c}{} &
        \multicolumn{4}{c}{} &
        \multicolumn{4}{c}{} &
        \colhead{} &
        \multicolumn{2}{r}{fit} &
        \multicolumn{2}{r}{err} &
        \multicolumn{2}{r}{fit} &
        \multicolumn{2}{r}{err} &
        \colhead{fit} & 
        \colhead{err}
    }
    \decimals
    \startdata
    1 & 05:46:06.01 & -00:09:32.70 & 0.36 & 0.08 & 4.60 & 1.03 & 5.47 & 4.62 & 59 & 5.26 & 1.22 & 4.43 & 1.07 & 58 & 53 \\
2 & 05:46:07.26 & -00:13:30.27 & 9.48 & 0.16 & 12.22 & 0.33 & 1.85 & 1.51 & 71 & 0.84 & 0.09 & 0.74 & 0.08 & 65 & 66 \\
3 & 05:46:07.33 & -00:13:43.49 & 31.37 & 0.16 & 36.88 & 0.31 & 1.78 & 1.43 & 70 & 0.68 & 0.03 & 0.56 & 0.03 & 57 & 12 \\
4 & 05:46:07.51 & -00:13:54.79 & 0.45 & 0.16 & 1.36 & 0.63 & 2.98 & 2.18 & 162 & 2.67 & 1.19 & 1.42 & 1.04 & 162 & 42 \\
5 & 05:46:07.53 & -00:11:49.22 & 0.97 & 0.14 & 5.25 & 0.90 & 4.34 & 2.69 & 152 & 4.14 & 0.74 & 2.14 & 0.49 & 153 & 11 \\
6 & 05:46:07.73 & -00:12:21.27 & 14.37 & 0.17 & 21.73 & 0.38 & 1.98 & 1.66 & 83 & 1.15 & 0.06 & 0.94 & 0.06 & 110 & 12 \\
7 & 05:46:07.84 & -00:09:59.61 & 6.45 & 0.11 & 7.59 & 0.21 & 1.73 & 1.36 & 65 & 0.81 & 0.07 & 0.36 & 0.10 & 62 & 8 \\
8 & 05:46:07.86 & -00:10:01.33 & 2.74 & 0.11 & 3.44 & 0.23 & 1.75 & 1.43 & 63 & 0.88 & 0.17 & 0.56 & 0.23 & 57 & 48 \\
9 & 05:46:08.42 & -00:10:01.03 & 0.86 & 0.09 & 5.53 & 0.69 & 4.75 & 2.70 & 39 & 4.52 & 0.60 & 2.33 & 0.34 & 38 & 8 \\
10 & 05:46:08.49 & -00:10:03.10 & 8.13 & 0.10 & 7.55 & 0.17 & 1.44 & 1.28 & 83 & -1.00 & -1.00 & -1.00 & -1.00 & -1 & -1 \\
11 & 05:46:08.92 & -00:09:56.11 & 2.07 & 0.11 & 2.28 & 0.20 & 1.58 & 1.39 & 74 & 0.51 & 0.31 & 0.36 & 0.21 & 129 & 73 \\
12 & 05:46:10.04 & -00:12:16.83 & 39.04 & 0.15 & 40.36 & 0.28 & 1.66 & 1.35 & 72 & 0.31 & 0.03 & 0.19 & 0.09 & 165 & 24 \\
13 & 05:46:13.13 & -00:06:04.94 & 9.41 & 0.14 & 11.18 & 0.27 & 1.66 & 1.38 & 67 & 0.71 & 0.07 & 0.50 & 0.08 & 60 & 19 \\
14 & 05:46:14.20 & -00:05:26.71 & 0.51 & 0.13 & 0.54 & 0.24 & 1.70 & 1.18 & 80 & -1.00 & -1.00 & -1.00 & -1.00 & -1 & -1 \\
15 & 05:46:27.91 & -00:00:52.11 & 65.62 & 0.14 & 73.72 & 0.27 & 1.66 & 1.41 & 74 & 0.52 & 0.02 & 0.49 & 0.02 & 43 & 26 \\
16 & 05:46:28.34 & +00:19:49.18 & 1.47 & 0.14 & 1.79 & 0.28 & 1.82 & 1.39 & 78 & 0.93 & 0.36 & 0.42 & 0.27 & 77 & 82 \\
17 & 05:46:28.61 & +00:20:58.08 & 0.50 & 0.13 & 0.49 & 0.23 & 1.68 & 1.21 & 127 & -1.00 & 1.59 & -1.00 & 0.60 & -1 & -1 \\
18 & 05:46:30.91 & -00:02:35.07 & 7.55 & 0.15 & 17.54 & 0.47 & 2.46 & 1.97 & 64 & 1.89 & 0.07 & 1.45 & 0.06 & 58 & 7 \\
19 & 05:46:31.09 & -00:02:32.95 & 16.15 & 0.15 & 25.05 & 0.35 & 1.86 & 1.74 & 152 & 1.31 & 0.03 & 0.73 & 0.05 & 160 & 5 \\
20 & 05:46:43.12 & +00:00:52.47 & 1.70 & 0.13 & 2.40 & 0.28 & 1.76 & 1.62 & 2 & 1.15 & 0.33 & 0.56 & 0.42 & 169 & 36 \\
21 & 05:46:46.52 & +00:00:16.09 & 1.00 & 0.12 & 1.04 & 0.22 & 1.53 & 1.37 & 64 & -1.00 & 1.05 & -1.00 & 0.50 & -1 & -1 \\
22 & 05:46:47.03 & +00:00:27.20 & 1.96 & 0.13 & 3.55 & 0.34 & 2.21 & 1.65 & 110 & 1.69 & 0.25 & 0.83 & 0.30 & 118 & 13 \\
23 & 05:46:47.43 & +00:00:23.24 & 1.10 & 0.09 & 12.48 & 1.10 & 5.31 & 4.32 & 38 & 5.11 & 0.47 & 4.09 & 0.39 & 36 & 22 \\
24 & 05:46:47.51 & +00:00:29.50 & 0.85 & 0.10 & 7.75 & 1.04 & 5.55 & 3.32 & 10 & 5.38 & 0.75 & 2.97 & 0.45 & 9 & 9 \\
25 & 05:46:47.69 & +00:00:25.02 & 5.38 & 0.13 & 7.22 & 0.27 & 1.81 & 1.50 & 54 & 1.01 & 0.11 & 0.64 & 0.13 & 37 & 15 \\
26 & 05:46:47.97 & +00:01:41.80 & 1.10 & 0.13 & 2.55 & 0.42 & 2.67 & 1.68 & 38 & 2.24 & 0.48 & 0.99 & 0.43 & 34 & 15 \\
27 & 05:46:57.30 & +00:23:57.94 & 3.39 & 0.16 & 4.25 & 0.33 & 1.67 & 1.42 & 72 & 0.83 & 0.21 & 0.55 & 0.34 & 61 & 72 \\
28 & 05:47:00.92 & +00:26:21.98 & 2.65 & 0.16 & 9.34 & 0.71 & 2.78 & 2.36 & 161 & 2.46 & 0.21 & 1.87 & 0.19 & 163 & 15 \\
29 & 05:47:01.31 & +00:26:23.09 & 4.91 & 0.17 & 6.96 & 0.37 & 1.78 & 1.48 & 93 & 1.05 & 0.12 & 0.72 & 0.13 & 99 & 21 \\
30 & 05:47:10.61 & +00:21:13.78 & 12.17 & 0.16 & 16.31 & 0.35 & 1.75 & 1.49 & 84 & 0.93 & 0.06 & 0.70 & 0.06 & 94 & 12 \\
31 & 05:47:15.95 & +00:21:22.89 & 2.02 & 0.13 & 2.70 & 0.28 & 1.80 & 1.55 & 89 & 0.92 & 0.29 & 0.75 & 0.51 & 111 & 77 \\
32 & 05:47:24.84 & +00:20:58.98 & 16.42 & 0.15 & 51.09 & 0.60 & 2.77 & 2.40 & 134 & 2.39 & 0.04 & 1.84 & 0.04 & 144 & 3 \\
33 & 05:47:32.45 & +00:20:21.60 & 5.56 & 0.15 & 7.19 & 0.30 & 1.74 & 1.50 & 99 & 0.91 & 0.14 & 0.60 & 0.20 & 129 & 21 \\
34 & 05:47:36.56 & +00:20:05.89 & 8.36 & 0.15 & 10.28 & 0.29 & 1.65 & 1.51 & 79 & 0.75 & 0.07 & 0.58 & 0.12 & 173 & 35 \\
    \enddata
    \tablenotetext{a}{Properties of the Gaussian fit to the ALMA emission: peak flux, integrated flux, major and minor axes of the FWHM, and position angle of the FWHM.}
    \tablenotetext{b}{Properties of the deconvolved Gaussian fit: major and minor axes of the FWHM, and position angle of the FWHM. Unresolved sources are indicated by values of -1.}
    \end{deluxetable*}
\end{longrotatetable}

\section{Detections} 
\label{sec:detections}

We detect a total of 34 continuum sources across 19 individual ALMA pointings. Table \ref{tab:observed-properties} lists each of the continuum sources with a running index number, location information based on the center of the two-dimensional Gaussian fit and additional associated statistics. These include the peak emission, the total flux, the major and minor axes, as well as the position angle. We also show the deconvolved values (and their uncertainties) for the size, and positional angle; with a -1 indicator if partially or fully unresolved.

\subsection{Protostellar Associations}
\label{sec:associations}

As introduced in Section \ref{sec:target-selection}, \citetalias{Nutter2007} identified 73 starless core candidates in their re-analysis of all SCUBA archival data in the Orion B North region. Since this time, there have been many studies of protostellar sources in the Orion molecular cloud complex, which have revealed many previously unknown protostars. Our observations will easily detect protostellar sources, and due to the statistical nature of our analysis, an accurate measure of the starless core population is needed. We have examined a number of different catalogs to check for any protostellar sources, which can be associated with our 34 ALMA detections. Table \ref{tab:detections-associations} lists each of our 34 ALMA detections along with each nearest protostellar source in all of the catalogs studied.

In summary, all but five of our ALMA detections are directly associated with a protostellar source, with one of those showing signs of a protostellar outflow in the CO data (see Section \ref{sec:alma-co-data} for details). The results from our protostellar catalog searches, as they pertain to the 34 ALMA detections are detailed in the following sections.

We perform an analogous check for the original starless core population identified by \citetalias{Nutter2007} for use in our statistical analysis (see Section \ref{sec:associations-summary} for details).

\subsubsection{Associations with Spitzer YSOs}
\label{sec:associations-spitzer}

We first search for the nearest Spitzer YSO in the \citet{Megeath2012} catalog. The separations found with this catalog fall into two main domains, those with very small separations ($<$2\arcsec), and those with larger separations ($>$13\arcsec). There are three such detections which lie in moderate separation range, source 23, 24, and 27, with a separation of 4.5\arcsec, 5.3\arcsec, and 7.8\arcsec to the nearest Spitzer YSO, respectively. Of these three Spitzer YSOs, the two nearest ALMA sources 23 and 24 are already more directly associated with other ALMA detections. The additional catalogues examined later in this section provided clarity on the protostellar/starless classification for these three detections.

In this study, we classify separations of less than 2\arcsec to the nearest Spitzer YSO to be coincident detections. In total, we find that 27 of our ALMA detections are found in the Spitzer catalog and we subsequently classify these detections as protostellar in nature.

\subsubsection{Associations with ALMA-based Catalogs}
\label{sec:associations-almabased}

We compared our detections to those found in \citet{Tobin2020}, The VLA/ALMA Nascent Disk and Multiplicity (VANDAM) Survey of Orion Protostars. This survey population was drawn from the Herschel Orion Protostellar Survey (HOPS) \citep[specifically][]{Fischer2010, Stutz2013, Furlan2016}, where all Class 0, Class 1 and Flat Spectrum protostars (with additional constraints) were selected for observations with ALMA Band 7 (0.87~mm), at a resolution of 0.1\arcsec (40~au), and with the VLA at 9~mm at a resolution of 0.08\arcsec (32~au). The VANDAM study observed a total of 328 protostars with ALMA, and only 148 protostars with the VLA \citep{Tobin2020}, so we choose to perform our association verification with the published ALMA catalog. Additional visual checks with both catalogs were conducted to ensure that associations in the Orion B North region were not missed. Compared to our ALMA data, the VANDAM observations have a higher angular resolution and reduced ability to recover flux on larger angular scales, hence VANDAM is expected to only detect protostellar sources.

Five ALMA detections do not lie within areas observed with VANDAM, and as such have no associations listed in the catalog. Of the 27 detections lying within VANDAM coverage, 25 have direct correspondences with VANDAM protostars, each with a separation of less than 0.50\arcsec. The remaining two detections, source 23 and 24, lie within the area observed by VANDAM, but no protostellar emission is seen by VANDAM and therefore should be starless.

Additionally, we compare our detections to the 19 sources analyzed by \citet{Dutta2020} in the Orion B region, through the ALMA Survey of Orion Planck Galactic Cold Clumps (ALMASOP). These are ALMA Band 6 (1.3~mm) observations, with a resolution of 0.35\arcsec (140~au). Due to the ALMASOP survey work covering the entire Orion complex, only nine of our detections are found in the ALMASOP survey, with separations less than approximately 0.5\arcsec, eight of which are classified as protostellar. These same eight sources are also identified as protostars by the VANDAM survey \citep{Tobin2020}. The one remaining ALMASOP associated detection, G205.46-14.56M3, is classified as starless by ALMASOP, in agreement with our final classification of source 1.

\movetabledown=12mm
\begin{longrotatetable}
    \tabletypesize{\scriptsize}
\begin{deluxetable*}{clDlDlDlDlDlDlDlD}
    \tablecaption{ALMA Detections and Nearest Catalog Objects \label{tab:detections-associations}}
    \tablehead{
        \colhead{Src} &
        \multicolumn{3}{c}{Spitzer} &
        \multicolumn{3}{c}{VANDAM} &
        \multicolumn{3}{c}{VISTA} &
        \multicolumn{3}{c}{SESNA} &
        \multicolumn{3}{c}{WISE} &
        \multicolumn{3}{c}{HGBS} &
        \multicolumn{3}{c}{APEX} &
        \multicolumn{3}{c}{ALMASOP} \\[-0.3cm]
        \colhead{} &
        \multicolumn{3}{c}{(1)} &
        \multicolumn{3}{c}{(2)} &
        \multicolumn{3}{c}{(3)} &
        \multicolumn{3}{c}{(4)} &
        \multicolumn{3}{c}{(5)} &
        \multicolumn{3}{c}{(6)} &
        \multicolumn{3}{c}{(7)} &
        \multicolumn{3}{c}{(8)} \\[-0.2cm]
        \colhead{\#} &
        \colhead{Name} & \multicolumn{2}{r}{($\arcsec$)} & 
        \colhead{Name} & \multicolumn{2}{r}{($\arcsec$)} & 
        \colhead{Name} & \multicolumn{2}{r}{($\arcsec$)} & 
        \colhead{Name} & \multicolumn{2}{r}{($\arcsec$)} & 
        \colhead{Name} & \multicolumn{2}{r}{($\arcsec$)} & 
        \colhead{Name} & \multicolumn{2}{r}{($\arcsec$)} & 
        \colhead{Name} & \multicolumn{2}{r}{($\arcsec$)} & 
        \colhead{Name} & \multicolumn{2}{r}{($\arcsec$)} 
    }
    \decimals
    \startdata
    1 & 3170 & 26.7 & HOPS-387-A & 38.5 & 129 & 40.3 & J054607.76-000937.7 & 26.9 & J054607.86-001000.8  & 39.8 & 1025 & 43.0 & 91 & 169.8 & G205.46-14.56M2 & 38.5 \\
    2 & 3161 & 0.5 & HOPS-358-A & 0.1 & 132 & 7.2 & J054607.26-001330.0 & 0.3 & J054604.78-001416.6  & 59.3 & 1021 & 3.6 & 91 & 69.8 & G205.46-14.56S1 & 0.2 \\
    3 & 3161 & 13.7 & HOPS-358-B & 0.2 & 132 & 20.5 & J054607.26-001330.0 & 13.5 & J054604.78-001416.6  & 50.5 & 1021 & 9.9 & 91 & 82.9 & G205.46-14.56S1 & 0.2 \\
    4 & 3161 & 25.2 & HOPS-358-B & 11.6 & 132 & 32.0 & J054607.26-001330.0 & 25.1 & J054604.78-001416.6  & 46.4 & 1021 & 21.5 & 91 & 94.1 & G205.46-14.56S1 & 11.6 \\
    5 & 3166 & 72.5 & HOPS-401 & 32.2 & 14 & 9.2 & J054607.88-001156.9 & 9.3 & J054607.86-001000.8  & 108.5 & 1024 & 5.8 & 91 & 31.6 & G205.46-14.56N2 & 32.2 \\
    6 & 3163 & 62.1 & HOPS-401 & 0.1 & 14 & 24.6 & J054607.88-001156.9 & 24.5 & J054604.78-001416.6  & 123.6 & 1024 & 26.6 & 91 & 1.3 & G205.46-14.56N2 & 0.1 \\
    7 & 3168 & 1.3 & HOPS-387-A & 0.2 & 129 & 2.1 & J054607.85-001001.1 & 1.5 & J054607.86-001000.8  & 1.3 & 1025 & 4.8 & 91 & 140.9 & G205.46-14.56M2 & 0.1 \\
    8 & 3168 & 0.4 & HOPS-387-B & 0.1 & 129 & 0.4 & J054607.85-001001.1 & 0.2 & J054607.86-001000.8  & 0.6 & 1025 & 4.5 & 91 & 139.2 & G205.46-14.56M2 & 1.7 \\
    9 & 3167 & 1.9 & HOPS-386-B & 0.5 & 129 & 8.5 & J054608.47-001002.9 & 2.0 & J054607.86-001000.8  & 8.3 & 1025 & 4.1 & 91 & 137.8 & G205.46-14.56M2 & 0.5 \\
    10 & 3167 & 0.5 & HOPS-386-A & 0.1 & 129 & 9.6 & J054608.47-001002.9 & 0.4 & J054607.86-001000.8  & 9.7 & 1025 & 5.5 & 91 & 135.6 & G205.46-14.56M2 & 2.8 \\
    11 & 3169 & 0.2 & HOPS-386-C & 0.1 & 129 & 17.0 & J054608.94-000956.0 & 0.2 & J054607.86-001000.8  & 16.6 & 1025 & 12.7 & 91 & 141.6 & G205.46-14.56M2 & 8.7 \\
    12 & 3163 & 77.9 & HOPS-402 & 0.0 & 14 & 37.9 & J054607.88-001156.9 & 37.9 & J054607.86-001000.8  & 139.8 & 1032 & 1.5 & 91 & 0.9 & G205.46-14.56N1 & 0.1 \\
    13 & 3180 & 0.4 & HOPS-388 & 0.2 & 96 & 32.2 & J054613.14-000604.8 & 0.2 & J054613.13-000604.8  & 0.2 & 1039 & 2.0 & 92 & 211.2 & G205.46-14.56M2 & 245.9 \\
    14 & 3183 & 0.2 & HOPS-320 & 0.1 & 157 & 0.5 & J054614.22-000526.5 & 0.3 & J054613.13-000604.8  & 41.4 & 1041 & 10.9 & 92 & 184.1 & G205.46-14.56M2 & 287.2 \\
    15 & 3195 & 95.8 & HOPS-403 & 0.1 & 178 & 95.5 & J054629.55-000135.1 & 49.5 & J054618.50-000017.8  & 145.2 & 1091 & 1.8 & 93 & 3.0 & G205.46-14.56M2 & 621.4 \\
    16 & 3338 & 0.4 & HOPS-331 & 0.2 & 120 & 200.6 & J054628.34+001949.3 & 0.2 & J054633.31+002255.5  & 200.7 & 1095 & 21.4 & 302 & 22.2 & G205.46-14.56M2 & 1814.4 \\
    17 & 3354 & 0.5 & HOPS-331 & 68.9 & 120 & 136.9 & J054628.59+002058.1 & 0.2 & J054633.31+002255.5  & 137.0 & 1095 & 90.4 & 302 & 91.2 & G205.46-14.56M2 & 1883.1 \\
    18 & 3195 & 160.9 & HOPS-373-B & 0.2 & 177 & 4.2 & J054629.55-000135.1 & 63.4 & J054618.50-000017.8  & 231.2 & 1108 & 0.8 & 93 & 111.8 & G205.46-14.56M2 & 558.7 \\
    19 & 3195 & 158.3 & HOPS-373-A & 0.3 & 177 & 7.1 & J054629.55-000135.1 & 62.3 & J054618.50-000017.8  & 232.0 & 1108 & 3.8 & 93 & 111.0 & G205.46-14.56M2 & 561.9 \\
    20 & 3208 & 0.1 & HOPS-363-A & 0.1 & 179 & 63.5 & J054643.11+000052.3 & 0.3 & J054643.12+000052.2  & 0.2 & 1152 & 5.5 & 93 & 254.0 & G205.46-14.56M2 & 835.0 \\
    21 & 3199 & 0.3 & HOPS-322 & 0.1 & 179 & 13.4 & J054646.51+000016.0 & 0.2 & J054643.12+000052.2  & 62.3 & 1168 & 15.7 & 93 & 257.1 & G205.46-14.56M2 & 840.5 \\
    22 & 3202 & 0.2 & HOPS-389-A & 0.1 & 179 & 0.0 & J054647.01+000026.9 & 0.2 & J054643.12+000052.2  & 63.5 & 1168 & 4.7 & 93 & 262.2 & G205.46-14.56M2 & 853.8 \\
    23 & 3201 & 4.5 & HOPS-323-B & 4.0 & 179 & 7.3 & J054647.69+000025.2 & 4.5 & J054643.12+000052.2  & 70.7 & 1168 & 3.2 & 93 & 255.6 & G205.46-14.56M2 & 855.1 \\
    24 & 3201 & 4.9 & HOPS-323-A & 5.0 & 179 & 7.9 & J054647.69+000025.2 & 5.0 & J054643.12+000052.2  & 69.7 & 1168 & 4.6 & 93 & 260.2 & G205.46-14.56M2 & 860.6 \\
    25 & 3201 & 0.3 & HOPS-323-A & 0.3 & 179 & 10.3 & J054647.69+000025.2 & 0.2 & J054643.12+000052.2  & 73.6 & 1168 & 5.6 & 93 & 255.1 & G205.46-14.56M2 & 859.1 \\
    26 & 3215 & 0.4 & HOPS-389-B & 75.2 & 179 & 76.0 & J054647.96+000141.8 & 0.2 & J054643.12+000052.2  & 88.0 & 1168 & 76.4 & 93 & 321.9 & G205.46-14.56M2 & 919.3 \\
    27 & 3398 & 7.8 & HOPS-338-B & 0.2 & 165 & 40.6 & J054657.35+002350.4 & 7.6 & J054703.69+002328.8  & 100.0 & 1202 & 2.0 & 96 & 124.1 & G205.46-14.56M2 & 2166.3 \\
    28 & 3409 & 1.2 & HOPS-341 & 0.5 & 126 & 38.3 & J054701.07+002544.1 & 38.0 & J054703.69+002328.8  & 178.1 & 1202 & 155.7 & 96 & 277.0 & G205.46-14.56M2 & 2320.1 \\
    29 & 3408 & 1.7 & HOPS-340 & 0.1 & 126 & 39.5 & J054701.07+002544.1 & 39.2 & J054703.69+002328.8  & 177.8 & 1202 & 159.0 & 96 & 280.6 & G205.46-14.56M2 & 2323.2 \\
    30 & 3359 & 0.4 & HOPS-365 & 0.1 & 103 & 1.8 & J054710.62+002113.8 & 0.1 & J054710.61+002114.0  & 0.3 & 1258 & 1.5 & 96 & 266.7 & G205.46-14.56M2 & 2093.5 \\
    31 & 3361 & 16.0 & HOPS-347 & 0.2 & 134 & 16.3 & J054714.89+002118.8 & 16.3 & J054712.90+002206.6  & 63.0 & 1285 & 1.6 & 96 & 343.8 & G205.46-14.56M2 & 2138.3 \\
    32 & 3356 & 1.0 & HOPS-359 & 0.4 & 141 & 29.0 & J054722.90+002058.0 & 29.1 & J054727.74+002035.8  & 49.4 & 1312 & 1.4 & 303 & 311.5 & G205.46-14.56M2 & 2184.3 \\
    33 & 3344 & 0.3 & HOPS-390 & 0.1 & 167 & 11.5 & J054732.44+002021.7 & 0.2 & J054736.60+002005.9  & 64.2 & 1338 & 1.5 & 303 & 204.0 & G205.46-14.56M2 & 2215.5 \\
    34 & 3341 & 0.4 & HOPS-364-B & 0.1 & 167 & 74.5 & J054736.57+002006.0 & 0.3 & J054736.60+002005.9  & 0.7 & 1360 & 5.2 & 303 & 151.9 & G205.46-14.56M2 & 2238.4 \\
    \enddata
    \tablecomments{Catalog entry name and separation distance to each of the protostellar catalogs entries, to each of the ALMA detections.}
    \tablerefs{(1) Megeath2012; (2) Tobin2020; (3) Spezzi2015; (4) R. Gutermuth 2024, in preparation; (5) Marton2016; (6) Konyves2020; (7) Stutz2013; (8) Dutta2020}
    \end{deluxetable*}
\end{longrotatetable}

\subsubsection{Other Published Catalogs}
\label{sec:associations-othercatalogs}

The following catalogs were also parsed for potential protostellar matches: the Herschel Gould Belt Survey \citep{Konyves2020}, the aforementioned Herschel Orion Protostellar Survey (HOPS) \citep{Stutz2013, Furlan2016}, the VISTA Orion Mini-Survey \citep{Spezzi2015}, the Spitzer Extended Solar Neighborhood Archive (SESNA) \citep[][R. Gutermuth et al. 2024, in preparation]{Pokhrel2020} and the Wide-field Infrared Survey Explorer (WISE) all-sky catalog \citep{Marton2016}. No new protostellar candidates were identified based on these catalogues, but the associations that we find are included in Table \ref{tab:detections-associations} for completeness.

\subsection{ALMA CO Data}
\label{sec:alma-co-data}

\begin{figure*}
    \gridline{
    \fig{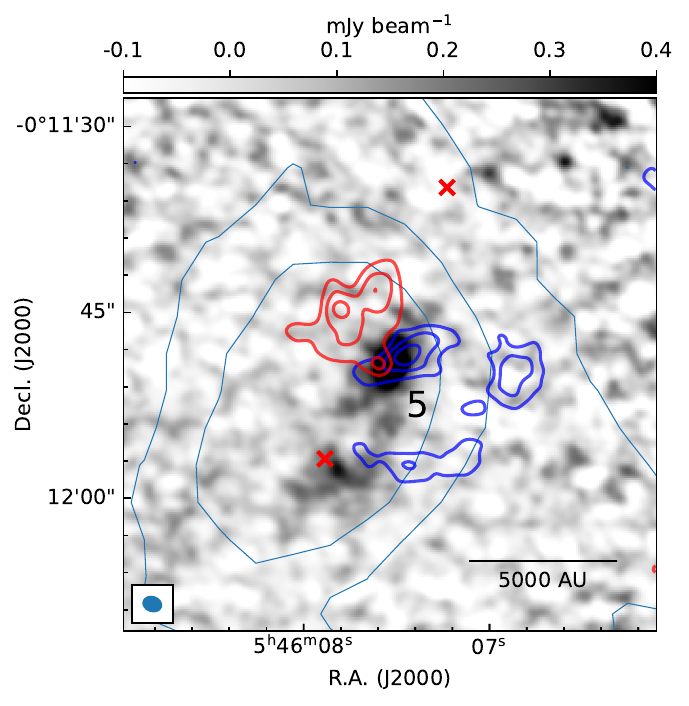}{0.5\textwidth}{}
    \fig{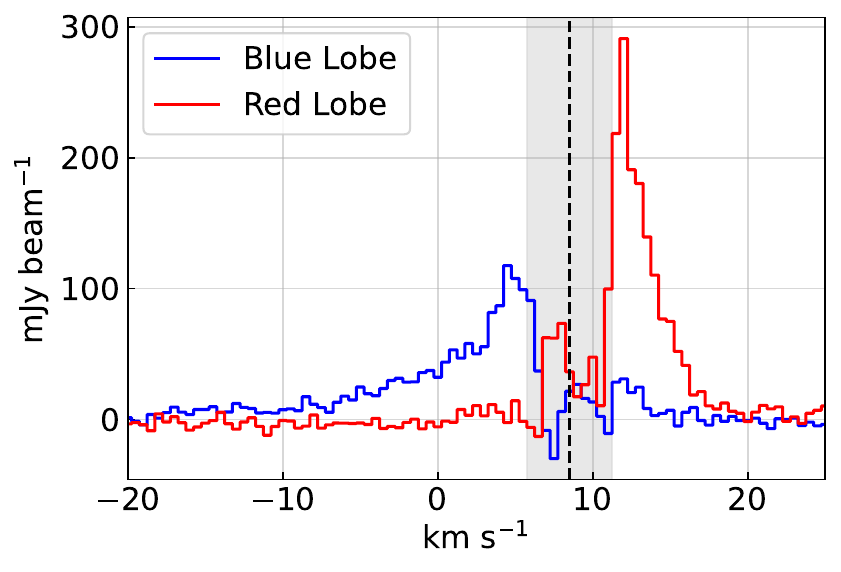}{0.5\textwidth}{}
    }
    \caption{
    Left: Zoom-in of the upper region of ALMA field BN-546074-01342Mosaic with source 5 shown in the center (full mosaic shown in Figure \ref{fig:Field1342}). The grayscale ranges linearly from -0.1~mJy~beam$^{-1}$ to 0.4~mJy~beam$^{-1}$, and the synthesized beam is given in the lower left corner within the frame. The light blue contours correspond to SCUBA2 850~$\mu$m emission at the corresponding levels in mJy~arcsec$^{-2}$: 0.15, 0.5, 1.0, 1.5, 3.0, 5.0. Protostellar sources in the field of view are plotted with red x markers. The blue and red contours represented velocity-shifted components of $^{12}$CO~$(1-0)$, with integrated velocity ranges of -10 to 5.75~km~s$^{-1}$ and 11.25 to 16~km~s$^{-1}$ respectively.
    The aforementioned contours are shown at 3$\sigma$, 5$\sigma$, 7$\sigma$  where the noise is 1$\sigma \sim$ 0.150~Jy~beam$^{-1}$~km~s$^{-1}$.
    Right: Spatially-averaged spectrum within the 3$\sigma$ contours of both red- and blue-shifted components as shown on the left. The grey shading indicates the velocity range excluded from the red and blue lobes, and the vertical black dashed line indicates the middle of this range. The lack of smaller-spacing data may be significantly filtering out emission within the velocity range indicated by the grey shaded region, near to the system velocity.
    \label{fig:Field1342-CO-Outflow}
    }
\end{figure*}

As introduced in Section \ref{sec:alma-data}, one spectral window of our ALMA observations was configured for observations of $^{12}$CO~$(1-0)$ emission at 115~GHz. We utilize the $^{12}$CO data to search for protostellar outflow signatures around the 5 ALMA detections which had no protostellar associations, as identified in Section \ref{sec:associations}. In summary, we find evidence for a protostellar outflow signature in only one of our imaged fields near source 5.

As shown in the left panel of Figure \ref{fig:Field1342-CO-Outflow}, source 5 shows red shifted and blue shifted CO emission perpendicular to the orientation of what appears to be a region of extended emission. The right panel of Figure \ref{fig:Field1342-CO-Outflow} shows that both the red- and blue-shifted emission peaks are located about 3.5~km~s$^{-1}$ offset from the central source velocity, and additional emission extends at least another 7~km~s$^{-1}$ in the red-shifted emission, and up to 26~km~s$^{-1}$ in the blue-shifted material, suggesting the presence of outflowing gas.

Source 5 also contains two nearby protostellar sources found through the VISTA survey \citep{Spezzi2015}. The upper source, 054607.227-001134.91, is approximately 15\arcsec away from our detection, and is labeled as a Class II source \citep{Spezzi2015}, and was previously found by \citet{Flaherty2008} and \citet{Fang2009}. The lower source, 054607.884-001156.83, is approximately 9\arcsec away from our detection, and is labeled as a potential Class III source \citep{Spezzi2015}.

Due to the absence of an associated infrared source at the location of source 5, this structure is likely a deeply embedded object, possibly a candidate first hydrostatic core. The first hydrostatic core stage is expected to have low-velocity molecular outflows at wide opening angles \citep{Fujishiro2020}, unlike the more collimated jets that accompany more mature protostellar sources. The velocity extent of the red-shifted material agrees well with the low-velocity (1-10~km~s$^{-1}$) molecular outflows predicted from simulations \citep{Machida2008}, however, the velocity extent of the blue-shifted material is outside the expected range. Further observations are needed to confirm the precise evolutionary status of this source.

The lower nearby protostellar source in Figure \ref{fig:Field1342-CO-Outflow}, 054607.884-001156.83 \citep{Spezzi2015}, appears to lie along a region of extended emission, to the southeast of our ALMA detection. The extended emission features in our ALMA observations are very low-level, only reaching slightly above 1$\sigma$. Nonetheless these low-level features may be indicative of streamer-like objects spanning thousands of au, recently seen in other star forming systems \citep[e.g.,][]{Pineda2020, Murillo2022, Pineda2023, Valdivia-Mena2023}, or an envelope associated with filamentary structure \citep{Tobin2010}. Recent studies have also indicated that the presence of ring-like structures around protostellar sources could be caused by magnetic flux removal as dense cores collapse during early stages of star formation \citep{Tokuda2023}, which may also be morphologically consistent with our observations.

In order to verify that this emission is indeed real and potentially associated with a streamer, deeper kinematic measurements tracing the velocity structure are required. We do not pursue further analysis of this potential extended emission here.

\subsection{Associations Summary}
\label{sec:associations-summary}

After consideration of the protostellar catalogs studied, with the addition of our ALMA observations in the CO, we classify sources 1, 4, 23, and 24 as likely being starless. The details of all of our associations are listed in Table \ref{tab:detections-associations}.

As some analysis is dependent on accurate classification of the original \citetalias{Nutter2007} dense core population (see Section \ref{sec:scuba-concentrations} for example), we perform the same check on that catalog for protostellar associations. We classify any protostellar object lying within 14\arcsec of the peak of the \citetalias{Nutter2007} core as associated, based on the beamsize of the SCUBA catalog. A total of 15 \citetalias{Nutter2007} SCUBA cores are re-classified as protostellar based on the catalogs studied, and will be used throughout the analysis when needed. In summary, we find that 58 fields are truly starless after verifying all ancillary data. This information is presented in Table \ref{tab:observation-noise-levels}, in the final column.

\subsection{Candidate Starless Core Detections}
\label{sec:starless-core-candidate}

\begin{figure*}
    \plotone{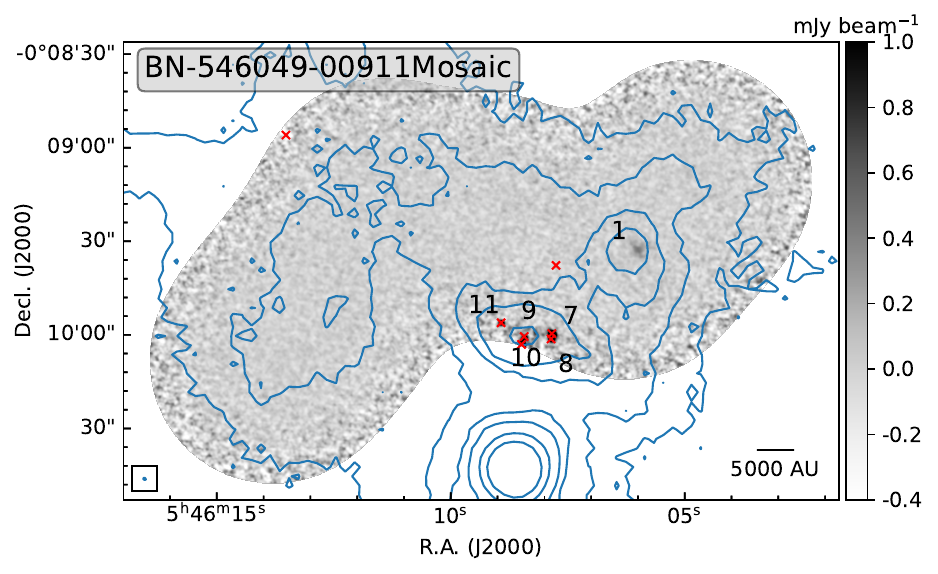}
    \caption{
    ALMA field BN-546049-00911Mosaic, with starless candidate source 1. The grayscale ranges linearly from -0.5~mJy~beam$^{-1}$ to 1.0~mJy~beam$^{-1}$. The blue contours correspond to SCUBA2 850~$\mu$m emission at the corresponding levels in mJy~arcsec$^{-2}$: 0.15, 0.5, 1.0, 1.5, 3.0, 5.0. All detections are labeled in black with their respective index number. Protostellar sources in the field of view are also plotted with red markers (see Table \ref{tab:detections-associations} for details). The synthesized beam is plotted in the lower left corner within a frame, along with a scalebar, indicating a linear distance of 5000~au, at the assumed distance to Orion of 419~pc.
    \label{fig:Field0911}
    }
\end{figure*}

Now we examine each of the four ALMA starless core detections in turn. Figure \ref{fig:Field0911} shows source 1, the only candidate starless core detection that is centered on a SCUBA-based target. This detection lies in the central portion of the SCUBA dense core BN-546063-00935 (\citetalias{Nutter2007}) and has a detection significance of 8 times the local rms noise. The area surrounding the peak position of the detection is quite extended, and forms a larger area of diffuse emission. As highlighted in Section \ref{sec:associations-almabased}, our source 1 is associated with the ALMASOP core G205.46-14.56M3 (G205-M3). G205-M3 is the only core in the ALMASOP dataset that shows signatures of fragmenting substructure at a scale of 1000~au \citep{Sahu2021}. The emission is resolved into two noticeable substructures with diameters of 1755~au and 820~au, and are approximately separated by a distance of 1200~au \citep{Sahu2021}. Due to our lower sensitivity and resolution compared to \citet{Sahu2021}, we are unable to resolve the two individual substructures within G205-M3. Nonetheless, \citet{Sahu2021} find that the enclosed masses, and their respective density profiles, are not consistent with a BE sphere-like model, and argue that this system likely represents an evolved starless state, just before the onset of star formation.

\begin{figure*}
    \epsscale{0.75}
    \plotone{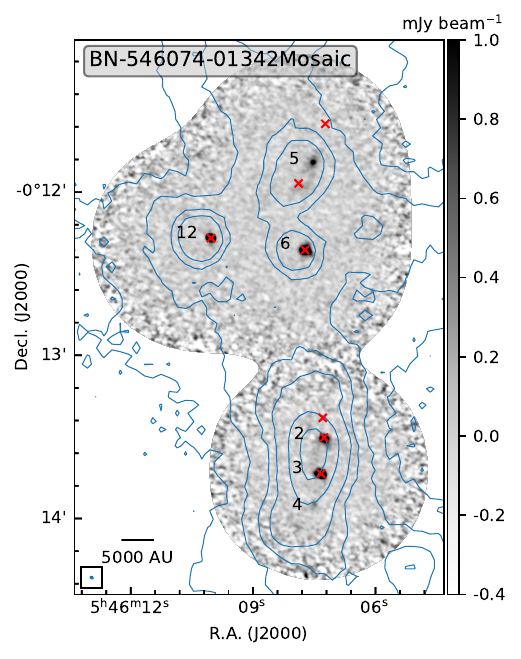}
    \caption{
    ALMA field BN-546074-01342Mosaic, with starless candidate source 4. See Figure \ref{fig:Field0911} for plotting conventions.
    \label{fig:Field1342}
    }
\end{figure*}

Figure \ref{fig:Field1342} shows a newly detected starless core candidate, source 4, and all other protostellar cores which lie in the same mosaic. This source is one of the faintest sources detected in our sample, with a peak flux measurement of 0.45~mJy~beam$^{-1}$, and has a detection significance of 7.4 times the local rms noise. This source is found to the south of multiple other protostellar sources in the SCUBA core BN-546074-01342, which have been positively identified as protostellar by \citet{Tobin2020} (see Table \ref{tab:detections-associations} for details).

It is striking that source 4 and the three previously identified protostellar sources appear regularly spaced along a line parallel to the elongation direction of the SCUBA emission, which is suggestive of formation via filament fragmentation \citep{Pineda2023}. This enhances the confidence of the source 4 detection despite its relatively low peak flux.

\begin{figure*}
    \plotone{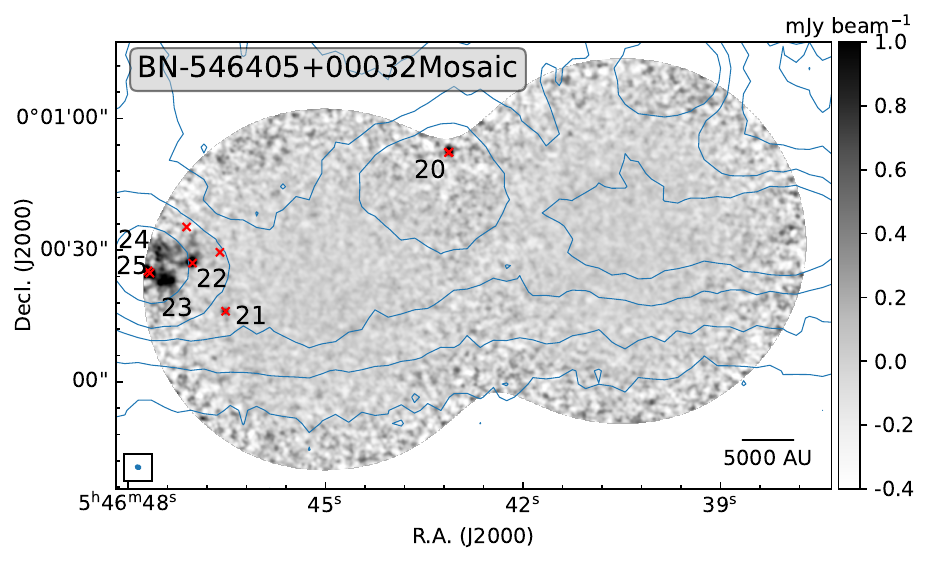}
    \caption{
    ALMA field BN-546405+00032Mosaic, with starless candidates sources 23 and 24. See Figure \ref{fig:Field0911} for plotting conventions.
    \label{fig:Field0032}
    }
\end{figure*}

Figure \ref{fig:Field0032} shows source 23 and 24, two additional newly detected fragments found within an area undergoing a large amount of fragmentation. These detections are located quite offset from the center of the \citetalias{Nutter2007} dense core BN-546450+00021. The larger area surrounding this dense core is home to a faint and complex emission structure, and more sensitive recent observations conducted by \citet{Kirk2016} indicate that there is a dense core associated with the positions of sources 23 and 24, with many individual fragmenting components\footnote{See Appendix \ref{sec:scuba-comparison} regarding the differences between the SCUBA and SCUBA-2 core catalogs for Orion B North.}. Note that the VANDAM protostellar survey (see Section \ref{sec:associations-almabased}) did cover the full region surrounding these detections with ALMA Band 7, but did not detect any emission for the locations associated with source 23 and 24. On the other hand, a recent 350GHz ACA Survey of 300 protostellar sources identified from the Herschel Orion Protostellar Survey was conducted by \citet{Federman2023}. These observations trace the protostellar flux at the envelope scale ($\le$8000~au), and find extended diffuse continuum detections at the locations of our source 23 and 24 \citep{Federman2023}. Thus, this region could very likely be at a moment in time where the transition from the starless core stage into the protostellar stage is being directly observed and currently ongoing.

In summary, out of the 4 starless core candidates identified, only one is found to be directly associated with one of the starless cores in the \citetalias{Nutter2007} sample. The other three all appear to be associated with fragmenting cores that already contain one of more protostellar sources.

\section{Derived Properties}
\label{sec:derived-properties}

Table \ref{tab:derived-properties} lists the physical properties of each of the continuum sources, including the mass estimate, effective radius, and number density. The effective radius is computed from a geometric mean of the semi-major and semi-minor axes of the deconvolved size, and if the source is unresolved, the synthesized beam is used in place, and written as an upper limit. 

\subsection{Mass Estimates}
\label{sec:mass-estimates}

\begin{deluxetable}{chcRcRc}
    \tablecaption{Physical Properties of Detections \label{tab:derived-properties}}
    \tablehead{
        \colhead{Src\tnm{a}} &
        \nocolhead{Source Name} &
        \colhead{Mass} & 
        \multicolumn{2}{c}{$R_{\text{eff}}$} & 
        \multicolumn{2}{c}{Number Density} \\ 
        \colhead{\#} &
        \nocolhead{} &
        \colhead{($M_{\odot}$)} &
        \nocolhead{Lim.} &
        \colhead{(au)} &
        \nocolhead{Lim.} & 
        \colhead{($\text{cm}^{-3}$)}
    }
    \colnumbers
    \startdata
    1 & BN-546049-00911MosaicA & 0.64 & ... & 1010 & ... & 1.9e+07 \\
    2 & BN-546074-01342MosaicD & 1.69 & ... & 170 & ... & 1.1e+10 \\
    3 & BN-546074-01342MosaicE & 5.10 & ... & 130 & ... & 7.2e+10 \\
    4 & BN-546074-01342MosaicF & 0.19 & ... & 410 & ... & 8.3e+07 \\
    5 & BN-546074-01342MosaicA & 0.72 & ... & 620 & ... & 9.1e+07 \\
    6 & BN-546074-01342MosaicC & 3.00 & ... & 220 & ... & 8.7e+09 \\
    7 & BN-546049-00911MosaicB & 1.05 & ... & 110 & ... & 2.1e+10 \\
    8 & BN-546049-00911MosaicC & 0.48 & ... & 150 & ... & 4.6e+09 \\
    9 & BN-546049-00911MosaicD & 0.76 & ... & 680 & ... & 7.4e+07 \\
    10 & BN-546049-00911MosaicE & 1.04 & < & 290 & > & 1.2e+09 \\
    11 & BN-546049-00911MosaicF & 0.32 & ... & 90 & ... & 1.3e+10 \\
    12 & BN-546074-01342MosaicB & 5.57 & ... & 50 & ... & 1.3e+12 \\
    13 & BN-546097-00552MosaicB & 1.55 & ... & 120 & ... & 2.4e+10 \\
    14 & BN-546097-00552MosaicA & 0.07 & < & 290 & > & 8.8e+07 \\
    15 & BN-546244-00001MosaicA & 10.18 & ... & 110 & ... & 2.6e+11 \\
    16 & BN-546287+02114MosaicA & 0.25 & ... & 130 & ... & 3.4e+09 \\
    17 & BN-546287+02114MosaicB & 0.07 & < & 290 & > & 8.0e+07 \\
    18 & BN-546244-00001MosaicC & 2.42 & ... & 350 & ... & 1.8e+09 \\
    19 & BN-546244-00001MosaicB & 3.46 & ... & 200 & ... & 1.2e+10 \\
    20 & BN-546405+00032MosaicA & 0.33 & ... & 170 & ... & 2.1e+09 \\
    21 & BN-546405+00032MosaicB & 0.14 & < & 290 & > & 1.7e+08 \\
    22 & BN-546405+00032MosaicC & 0.49 & ... & 250 & ... & 9.7e+08 \\
    23 & BN-546405+00032MosaicE & 1.72 & ... & 960 & ... & 5.9e+07 \\
    24 & BN-546405+00032MosaicF & 1.07 & ... & 840 & ... & 5.5e+07 \\
    25 & BN-546405+00032MosaicD & 1.00 & ... & 170 & ... & 6.3e+09 \\
    26 & BN-546499+00204A & 0.35 & ... & 310 & ... & 3.5e+08 \\
    27 & BN-546580+02426A & 0.59 & ... & 140 & ... & 6.3e+09 \\
    28 & BN-547014+02614B & 1.29 & ... & 450 & ... & 4.3e+08 \\
    29 & BN-547014+02614A & 0.96 & ... & 180 & ... & 4.8e+09 \\
    30 & BN-547103+02112A & 2.25 & ... & 170 & ... & 1.4e+10 \\
    31 & BN-547160+02123MosaicA & 0.37 & ... & 170 & ... & 2.1e+09 \\
    32 & BN-547252+02059MosaicA & 7.06 & ... & 440 & ... & 2.5e+09 \\
    33 & BN-547336+01902MosaicB & 0.99 & ... & 150 & ... & 8.1e+09 \\
    34 & BN-547336+01902MosaicA & 1.42 & ... & 140 & ... & 1.6e+10 \\
    \enddata
    \tablenotetext{a}{Running Index Number, the same as Table \ref{tab:observed-properties}.}
    \tablecomments{(3) and (5) show limit indicators for unresolved sources. For these sources, the synthesized beam has been used in-place for the size; the effective radius should be taken as an upper limit, while the number density should be taken as a lower limit.}
\end{deluxetable}

We estimate the mass of each continuum source using the standard equation:

\begin{equation}\label{eq:core-mass}
    M = 100 \frac{d^2 S_{\nu}}{B_{\nu}(T_D) \kappa_{\nu}} \ ,
\end{equation}

\noindent where $d$ is the distance, $S_{\nu}$ is the integrated flux at frequency $\nu$, $B_{\nu}$ is the Planck function at the dust temperature of $T_D$, and the factor of 100 represents the gas-to-dust ratio. We adopt a slightly updated value of distance, $d=419~\text{pc}$, consistent with newer estimates from \citet{Zucker2019}, along with a core temperature of $T=10~\text{K}$.

We use commonly adopted calculations of opacities, \citet{Ossenkopf1994}, specifically the OH5 model corresponding to thin ice mantles after $10^5$~years of coagulation at a gas density of $10^6$~cm$^{-3}$. Extending the model to an effective frequency of 106~GHz yields our choice in opacity of $\kappa_{\nu} = 0.23~\text{cm}^2~\text{g}^{-1}$. As mentioned in \citet{Dunham2016}, specific choices in opacity typically have uncertainties of factors $2-4$ \citep[see also ][]{Shirley2005,Shirley2011} along with additional uncertainties in the power-law index due to dependence on grain size \citep{Ricci2010a,Ricci2010,Tobin2013,Schnee2014,Testi2014}. Since the mass is inversely proportional to $\kappa$, reducing $\kappa$ to half the value we presently adopt \citep[as is used in][]{Motte1998} would increase masses by a factor of two.

We calculate the mean density of each source as

\begin{equation}\label{eq:number-density}
    n = \frac{3}{4 \pi \mu m_{\text{H}}} \frac{M}{R_{\text{eff}}^3} \ ,
\end{equation}

\noindent where $\mu = 2.80$ is the mean molecular weight per hydrogen molecule \citep{Kauffmann2008} and $m_{\text{H}}$ is the mass of a hydrogen atom.
Both the computed mass and number density estimates are given in Table \ref{tab:derived-properties}. The uncertainties are dominated by systematic effects (such as temperature, distance, dust opacity, etc.), rather than the statistical uncertainties associated with the Gaussian fits performed to characterize the size of the sources.

\subsection{ALMA Peak Flux}
\label{sec:alma-peak-flux}

\begin{figure}
    \plotone{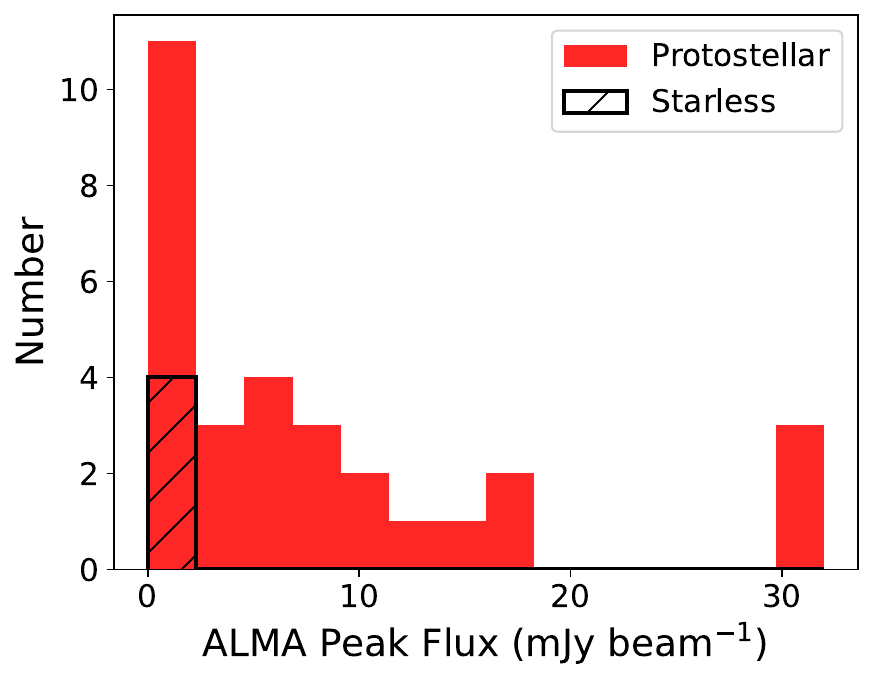}
    \caption{
    Distribution of ALMA detection peak fluxes. Detections with a peak flux greater than 30~mJy~beam$^{-1}$ are plotted in the final bin shown. We show both the protostellar ALMA detections in red (30 sources total), along with the candidate starless core detections in black (4 sources total).
    \label{fig:ALMA-peak-fluxes}
    }
\end{figure}

Figure \ref{fig:ALMA-peak-fluxes} shows the distribution of peak fluxes for both the protostellar and starless population of our ALMA detections. It is expected that the later stages of prestellar core evolution passes relatively quickly \citep{Jessop2000, Girichidis2014, Zamora-Aviles2014}, in which the stage at which the prestellar core is detectable. Compared to the development of a protostellar core, a starless core typically has a lower central density, and as a result, we should expect to see our prestellar cores have a low peak flux compared to the protostellar cores in our sample. All of our starless core detections notably lie in the lowest peak flux bin, suggesting that our observations agree with the overall picture of starless core evolution.

\subsection{SCUBA Core Concentration}
\label{sec:scuba-concentrations}

\begin{figure}
    \plotone{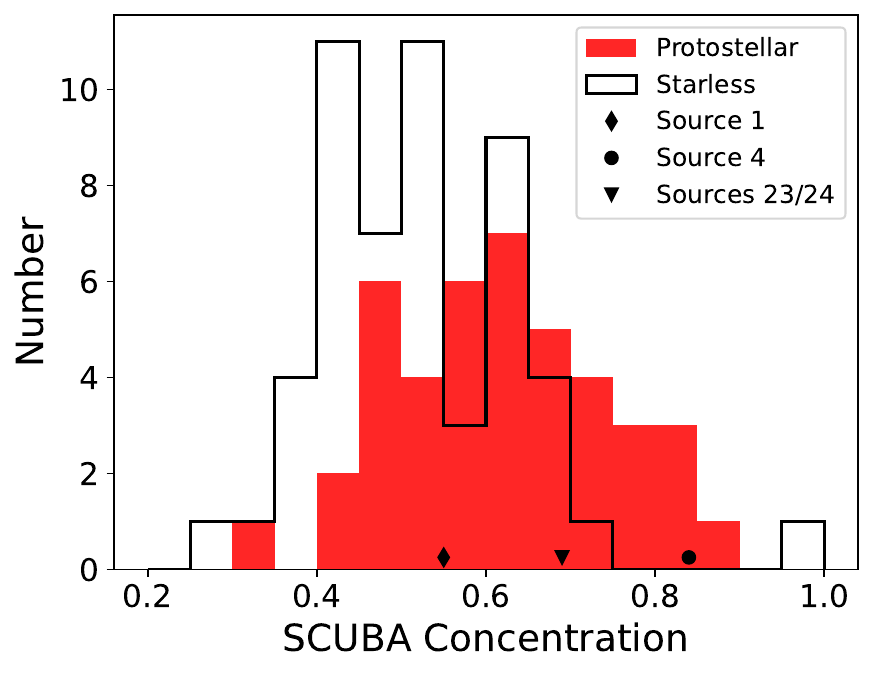}
    \caption{
    Distribution of core concentrations for dense cores in the Orion B North star forming region, computed from the re-classified \citetalias{Nutter2007} catalog (see Section \ref{sec:associations-summary} for details). See Figure \ref{fig:ALMA-peak-fluxes} for plotting conventions. Markers indicate the concentrations of the parents core of the associated labeled candidate starless core(s). We have excluded all values below a concentration of 0.2 (see Section \ref{sec:scuba-concentrations} for discussion regarding this choice).
    \label{fig:SCUBA-core-concentrations}
    }
\end{figure}

We measure the central concentration of dense cores from the \citetalias{Nutter2007} catalog, with the following definition:

\begin{equation}\label{eq:concentration}
    C = 1 - \frac{1.13 B^2 F_{\text{tot}}}{\pi R^2 F_{\text{pk}}} \ ,
\end{equation}

\noindent where $B$ is the beamsize, $F_{\text{tot}}$ is the integrated flux, $R$ is the effective radius, and $F_{\text{pk}}$ is the peak flux \citep{Johnstone2000}.

The importance of self-gravity alongside the internal pressures of dense core structure, set the possible collapse of these structures into the smaller scales, usually associated with protostars. Bonnor-Ebert spheres (BE Spheres) are a spherical, isothermal model for self-gravitating  pressure-confined objects \citep{Ebert1955, Bonnor1956}, and this simple model predicts a range of values in terms of central concentrations. It is expected that dense cores become more centrally concentrated as they evolve in time, with uniform spheres having a central concentration value of 0.33 while the central concentration value of 0.72 is the maximum allowed for a stable BE sphere \citep{Johnstone2000}. Any cores above such a value do not have equilibrium solutions, and therefore must undergo gravitational collapse. Dense cores that are high in central concentration are usually found to be protostellar in nature \citep[e.g.,][]{Johnstone2000,Kirk2006,Jorgensen2008}.

Figure \ref{fig:SCUBA-core-concentrations} shows the distribution of central concentration for all \citetalias{Nutter2007} cores in Orion B North. There are six starless dense cores that have concentrations lower than 0.33, the minimum value allowed for a uniform sphere. The \citetalias{Nutter2007} catalog for the Orion B North area, has a reported 1$\sigma$ rms of 16~mJy~beam$^{-1}$, and the analysis uses a detection threshold of $>5\sigma$ relative to the local background for a robust source classification. Due to the lack of precision published in the values for the peak flux and the integrated flux \citepalias{Nutter2007}, for values of emission near the lowest reported limit of 0.1~Jy~beam$^{-1}$, there exists a large amount of uncertainty. Since the central concentration uses both the measure of peak flux, as well as the total integrated flux, for the faintest cores, associated errors and uncertainties would propagate much more significantly. In this case, some reported values are found below the allowed minimum value of 0.33.\footnote{Additionally, for cores in close proximity to other cores in the catalog, \citetalias{Nutter2007} necessarily reduced the elliptical aperture size used to perform the 2D fit, to a level below the usually chosen 3$\sigma$ contour. This choice partitions the extended emission between the cores, and could also contribute to the overall uncertainty in the measurements reported.} The surprisingly high concentration value of 0.955 associated with a starless core (the most elongated core reported in the \citetalias{Nutter2007} sample) similarly appears to be influenced by the same reasons as above. In all cases, the associated SCUBA-2 dense cores do have concentrations above the minimum 0.33 \citep{Kirk2016} (see Appendix \ref{sec:scuba-comparison} for discussion).

Our starless core candidates lie in SCUBA dense cores with concentration values of 0.55 (source 1), 0.84 (source 4), and 0.69 (sources 23 and 24). While there is a range of concentrations for both the starless and protostellar population of cores, as seen in Figure \ref{fig:SCUBA-core-concentrations}, more starless cores are found at lower concentrations, as expected for starless core populations in which they are more likely found in areas of extended emission still undergoing mass accretion and gravitational collapse. We find that our starless core detections are either within the typical range for protostellar dense cores, or on the higher side for starless dense cores.

In the analysis of the Ophiuchus molecular cloud, \citet{Jorgensen2008} found four starless cores with a high degree of central concentration, contrary to the previous study in the Perseus molecular cloud, where no starless cores were found to have a central concentration higher than 0.6 \citep{Jorgensen2007}. \citet{Kirk2017} subsequently found no ALMA detections for SCUBA starless cores in Ophiuchus with concentrations below 0.6, but did detect all starless cores with higher concentrations than 0.6. Further evolved starless cores are expected to have a higher degree of central concentration as they collapse towards a protostellar state. All of our starless cores detections exhibit a high degree of central concentration, reinforcing this picture.

\section{Substructure and Fragmentation in Starless Cores}
\label{sec:substructure-fragmentation}

\subsection{Numerical Simulations and Synthetic Observations}
\label{sec:numerical-simulations}

Turbulent simulations predict that fragmentation within dense cores begins during the starless core phase \citep{Offner2012}. Interferometers like ALMA are uniquely suited to detecting these small and faint density peaks within starless cores, while simultaneously filtering out larger-scale emission structures. In this section, we perform the same approach as \citet{Dunham2016} and \citet{Kirk2017}, and compute the number of starless core detections predicted by the turbulent fragmentation model for our observed Orion B North core sample.

We use magneto-hydrodynamic simulations of isolated, collapsing starless cores, with an initial core mass of $0.4M_{\odot}$, using the ORION Adaptive Mesh Refinement (AMR) code base \citep{Li2021}, to generate self-consistent, time-dependent physical conditions. The simulation starts with a uniform density spherical core of gas, at a temperature of 10~K, with an initial number density of $1.6 \times10^5~\text{cm}^{-3}$ and uniform magnetic field in the $z$ direction \citep[e.g., similar to][]{Offner2017}. The core is surrounded by a warm, low density medium, with a temperature of 1000~K, and a density 100 times lower than the initial core density. At the starting time step in the simulation, the gas velocities in the core are perturbed with a turbulent random field, and once set in motion, are allowed to decay with no additional energy injection. ORION evolves the calculations until shortly after the formation of a first hydrostatic core, which is represented by a sink particle. For more in-depth details on the descriptions of the simulations, we refer to the reader to Section 5.1.1 in \citet{Dunham2016}.

To generate synthetic ALMA observations, we use the same turbulent simulation snapshots as in \citet{Dunham2016} and \citet{Kirk2017}. We take the total gas column density snapshots and convert them to total gas surface density, using the same mean molecular weight per free particle as in Equation \ref{eq:number-density}. We then derive the mass in each pixel, and compute the flux map using Equation \ref{eq:core-mass}.

Finally, we create the synthetic observations using the appropriate antenna configurations, on-source time, and distance to Orion B North to match our current ALMA observations. The synthetic observations were generated using CASA's \texttt{simalma} task, and re-imaged with CASA's \texttt{tclean} task to mimic all imaging parameters used in the physical observations (see Section \ref{sec:alma-data} for details). We choose the position of R.A. = 05:47:00 and Decl. = +00:05:00 for our simulated observations, which equates roughly to the center of our observational area in Orion B North. We use a total on source time of 270~s, with 190~s at the more compact configuration, and 80~s at the more extended configuration, to match our real observations. We use integration times of 2~s at an effective mean frequency of 106~GHz with a bandwidth of 6~GHz, include the default atmospheric noise model, and the same imaging cell size of 0.17\arcsec used to image the real observations.

\begin{figure*}
    \plotone{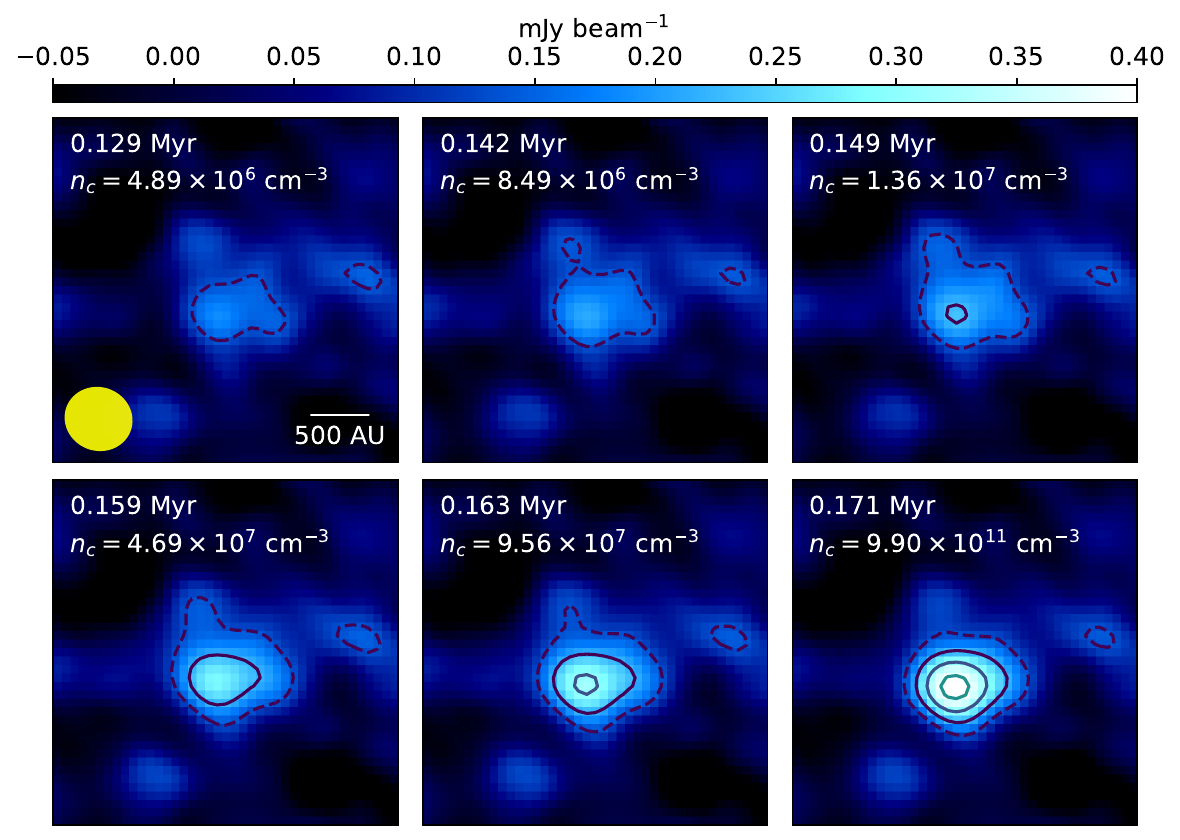}
    \caption{
    Synthetic ALMA 106 GHz observations of the $0.4M_{\odot}$ simulation, at six given timesteps, indicated in each panel along side the central density of the core. The synthesized beam is given in the first panel as a yellow ellipse in the bottom left corner. The dashed contour represents the 3$\sigma$ level. The solid contours start at a level of 5$\sigma$ and increase by 2$\sigma$, where 1$\sigma \sim$~0.045~mJy~beam$^{-1}$. We consider a robust detection to be a minimum of 5$\sigma$.
    \label{fig:synthetic_alma}
    }
\end{figure*}

Figure \ref{fig:synthetic_alma} shows six time steps (from 0.129~Myr to 0.171~Myr) of the synthetic ALMA observations of the $0.4M_{\odot}$ core simulation.

\subsection{Detecting Starless Cores}
\label{sec:detecting-starless-cores}

With the assumption of a continuous rate of star formation over the timescale at least as long as the core lifetimes, then similar to \citet{Dunham2016} and \citet{Kirk2017}, we can compute the expected number of detections as the following:

\begin{equation}\label{eq:num-of-detections}
    \text{Detections} > \frac{2}{3} \times N_{\text{total}} \times \left( 
    \frac{n_{\text{Detectable}}}{n_{\text{Limit}}}
    \right)^{-0.5} \ ,
\end{equation}

\noindent where $N_{\text{total}}$ is the number of starless cores observed, $n_{\text{Detectable}}$ is the core density at which our ALMA observations can detect the core, and $n_{\text{Limit}}$ is the observed lower limit of the mean core densities as observed at single-dish resolution (in this case, the \citetalias{Nutter2007} data).

\begin{figure}
    \plotone{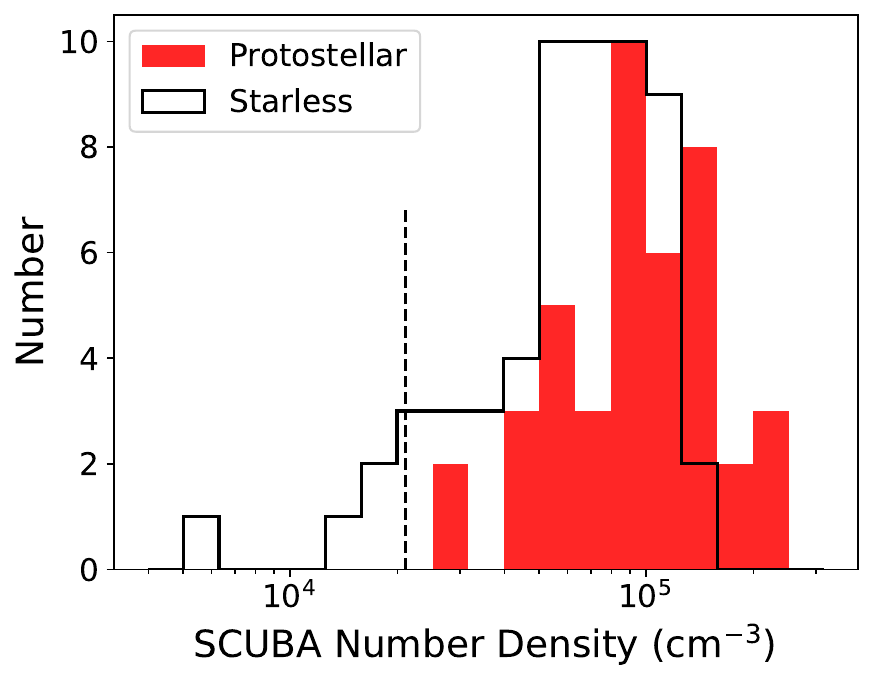}
    \caption{
    Distribution of \citetalias{Nutter2007} core number densities. The black histogram indicates cores classified as starless, while the red filled histogram indicates cores classified as protostellar in nature. The vertical black dashed line indicates the minimum detectable density, as computed in Section \ref{sec:detecting-starless-cores}.
    \label{fig:SCUBA-core-number-densities}
    }
\end{figure}

Figure \ref{fig:SCUBA-core-number-densities} shows the distribution of core number densities for all cores in the Orion B North star forming region as presented in \citetalias{Nutter2007}. The lowest density cores found in our population, near the left tail of the distribution (along with the one significant outlier) correspond to cores with atypical properties (highly elongated and very large cores), so we briefly explain our choice for the observed lower limit chosen for our calculations\footnote{The most atypical core, BN-547034+1950, is the same extremely high concentration core as detailed in Section \ref{sec:scuba-concentrations}.}. We compute the number density of a dense core whose size is equal to the beamsize (14\arcsec) and which lies at the sensitivity limit of the \citetalias{Nutter2007} dataset, to represent an object at the minimum detectable density. \citetalias{Nutter2007} classifies any object at 5$\sigma$ above the local noise level as a positive detection, and for the Orion B North area, the 1$\sigma$ rms noise is given as 16~mJy~beam$^{-1}$. This corresponds to a minimum detectable density of $2.11 \times10^4~\text{cm}^{-3}$. For comparison, our lowest density starless core observed has an estimated density of $5.68 \times10^3~\text{cm}^{-3}$, although as can be seen in Figure \ref{fig:SCUBA-core-number-densities}, only a small number of the starless cores have estimated densities below $2.11 \times10^4~\text{cm}^{-3}$. Our derived value of $2.11 \times10^4~\text{cm}^{-3}$ agrees well with the lower tail of the distribution as shown in Figure \ref{fig:SCUBA-core-number-densities}, and is representative of a typical core seen in our observations.

The central core peak in the simulations is only detected at a 5$\sigma$ level after 0.149~Myr, when the central core density reaches $1.36 \times 10^7~\text{cm}^{-3}$. We use this as the criteria for a robust detection, and we set the detectability limit as such.

In total, for the Orion B North region studied, we have $N_{\text{total}} = 58$~starless cores (as re-classified in Section \ref{sec:associations-summary}), with a minimum density of $n_{\text{Limit}} = 2.11 \times 10^4~\text{cm}^{-3}$, along with the detectability limit of our ALMA configuration $n_{\text{Detectable}} = 1.36 \times 10^7~\text{cm}^{-3}$. Using Equation \ref{eq:num-of-detections}, along with Poisson statistics, we predict a minimum of $2 \ (1.52) \pm 1$ starless cores that have enough substructure for a positive detection in our observations.

As shown in Figure \ref{fig:SCUBA-core-number-densities}, there exists a small number of starless cores that have estimated densities lower than our chosen density limit. For example, our lowest estimated density core with a density of $5.68\times10^3~\text{cm}^{-3}$ belongs to the same highly elongated core as mentioned earlier in this section. If we instead utilize this density in place of our chosen density limit, we would lower our expected number of starless core detections to 0.79. The results would remain unaffected with either difference in chosen detectability limit.

For completeness, we also consider the potential uncertainty in the total number of cores observed, for example, if the \citetalias{Nutter2007} catalogue over-segmented cores lying within filamentary structure. As described in Appendix \ref{sec:scuba-comparison}, we perform a careful visual comparison between the \citetalias{Nutter2007} core catalog with the core catalog of \citet{Kirk2016} of the same region using newer SCUBA-2 data and an independent core-identification technique. While the vast majority of the cores identified agree well, we note several instances where the two catalogs differ. Conservatively, we estimate that the true number of starless cores observed from the \citetalias{Nutter2007} catalog lies between 47 and 61. This range places the number of predicted detections between 1.23 and 1.60.

In addition, we note that various dust and gas tracers used for core identification can lead to differences in the boundaries of identified cores \citep[e.g.,][]{Ikeda2009}, leading to further uncertainties in the total number of cores observed, which could increase the range in true core numbers reported above.

Due to the re-classification of the original \citetalias{Nutter2007} SCUBA core catalog, only one of our total four starless core candidates lie in a truly starless dense core (see Section \ref{sec:associations-summary}). Thus, we show that this one starless core detection is consistent with the model predictions. Additionally, we also run the same synthetic observations on the simulation of a $4M_{\odot}$ starless core and find less than a factor of two difference in the detectability threshold despite the change in initial mass of an order of magnitude, suggesting a similar number of detections (1.23) (see Appendix \ref{sec:4M_simulations} for details).

As introduced in Section \ref{sec:intro}, a BE sphere-like model has a smooth and broad density profile, as driven by completely thermal evolution processes. Observations taken with an interferometer leads to filtering effects, which depending on the scale of emission, can impact the detectability of such objects. \citet{Dunham2016} demonstrated that structures generated from the collapse of turbulent cores should be detectable at a rate approximately 100 times higher than the BE sphere-like model for their ALMA observational setup of Chamaeleon I. Since our ALMA observations are very similar to those taken in both the Chamaeleon I ALMA study \citep{Dunham2016} and the Ophiuchus ALMA study \citep{Kirk2017}, we should also expect to see a lack of detections if the evolution and resulting density structure is similar to that of a BE sphere-like model. 

A single detection under the turbulent fragmentation model is not a large number for model testing, however, we note that under the BE sphere-like model, it would be very unlikely to get a single detection. With the combination of Orion B North and Ophiuchus \citep{Kirk2017}, we now have two studies which show that the fragmentation model serves to represent a clearer picture of the on-going evolution of starless core evolution.

\subsection{Evolutionary Lifetime Estimate}
\label{sec:starless-core-lifetimes}

Relative source counts are a popular way to estimate the lifetime of the earlier phases of star formation \citep[e.g.,][]{Beichman1986, Jessop2000, Kirk2005}. Assuming a continuous star formation rate, the ratio of lifetimes of earlier and later phase objects is directly reflected in the relative number counts. Many measures for nearby star forming regions currently exist with central densities between $\sim10^3$ and $\sim10^5$ \citep[see review by ][]{Ward-Thompson2007}. We can apply a similar procedure for our single ALMA detected starless core to obtain a crude estimate at slightly higher central densities, noting the inherently large uncertainty with only one starless core detected in our sample.

We adopt a reference Class I protostellar lifetime of 0.74~Myr from \citet{Konyves2015} derived from relative protostellar counts of Spitzer data of Orion B \citep{Megeath2012}. By comparing the 37 Class I cores found within the \citetalias{Nutter2007} survey footprint from the SESNA catalog (R. Gutermuth et al. 2024, in preparation) to our one ALMA starless core detection, we estimate the lifetime for ALMA detectability to be $2.0\times10^4$~yr.

\begin{figure}
    \plotone{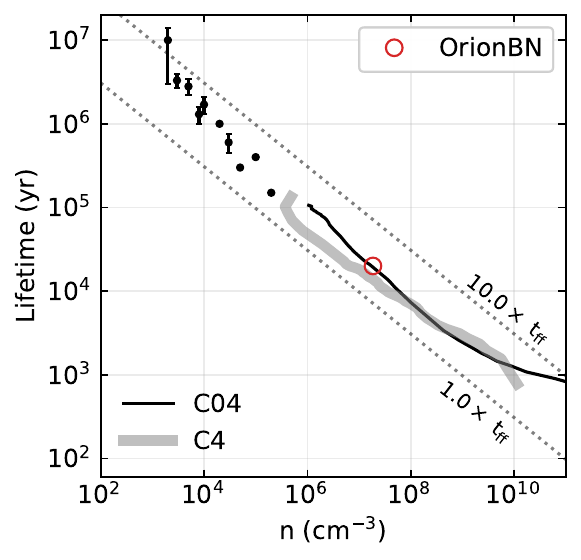}
    \caption{
    Lifetimes vs. central number densities for the $0.4M_{\odot}$ simulation (C04) and the $4.0M_{\odot}$ simulation (C4), as reproduced from Figure 10 in \citet{Dunham2016}. The lifetime at each given density is the amount of time it takes the simulation to evolve from that given timestep to the onset of the first hydrostatic core phase. The dashed lines indicate the free-fall lifetime ($t_{ff} = \sqrt{(3\pi)/(32G\rho)}$) and $10t_{ff}$. The data points show early sub-mm observations at smaller densities \citep[see ][]{Jessop2000, Kirk2005}. The red circular marker shows the statistical lifetime derived with the Orion BN ALMA observations.
    \label{fig:statistical lifetime}
    }
\end{figure}

Our starless core has an estimated central density of $1.86 \times 10^7~\text{cm}^{-3}$, implying a free-fall timescale of $7.12 \times 10^3~\text{yr}$. As shown in Figure \ref{fig:statistical lifetime}, our estimated ALMA-detected starless core lifetime is larger than this free-fall timescale by a factor of 2.80, however, we emphasize the large uncertainties associated with our estimate given that we only have one detection. We interpret our result as implying the core lifetime at a density of approximately $10^7~\text{cm}^{-3}$ is more consistent with a shorter timescale (i.e., closer to the free-fall time) rather than a longer timescale (i.e., closer to 10 times the free-fall time), and note that a larger sample size is necessary to get a firmer lifetime estimate. The shorter lifetime for our ALMA detection is consistent with the numerical simulations, that we analyze, where the lifetimes tend to lie closer to $t_{ff}$ than $10\times t_{ff}$. 

\section{Comparison to Chamaeleon I and Ophiuchus}
\label{sec:comparisons}

Through the analysis of the Chamaeleon I star forming region, \citet{Dunham2016} expected two starless core detections under the turbulent fragmentation model, but found no such detections in their observations. \citet{Dunham2016} proposed three arguments as to why no starless core detections were found in the Chamaeleon I cloud: the lack of continuous star formation in the cloud, that the assumption of the core lifetime proportionality to the free-fall time is not correct, or that the numerical simulations are not applicable. This is in direct contrast to the study performed by \citet{Kirk2017}, where the dense core population of the Ophiuchus star forming region was analyzed, and found to agree with those same fragmentation models with two tentative detections compared with two predicted.

With the addition of the Orion B North observations, we are now in a position to search for differences in the core populations per cloud, as a possible explanation for the lack of detections in Chamaeleon I. A basic virial analysis was performed by \citet{Tsitali2015} in which the core's self-gravity contribution is compared with the thermal and non-thermal pressure support within the core. From this analysis, \citet{Tsitali2015} showed that the population of dense cores in the Chamaeleon I cloud are gravitationally unbound, suggesting that star formation may not continue. 

A full virial analysis allows for more information regarding specific physical processes that may be guiding differences in star forming regions. Some additional virial terms are pressure support from cloud-wide turbulence, pressure support from the molecular cloud weight, and contributions from magnetic fields \citep[e.g.,][]{Pattle2015,Kirk2017a,Kerr2019}. 

We carry out both the simple Jeans analysis and a fuller virial analysis including the larger-scale external pressure binding terms, to investigate the difference in the predicted number of detections between the star-forming clouds studied. Since our goal is to compare the core populations between clouds, we adopt comparable datasets for all three clouds; unfortunately no such uniform data was available for the magnetic field strength, so we exclude this term. We follow with descriptions of the adopted comparable datasets as they pertain to both the Jeans analysis and the fuller virial analysis. 

\subsection{Datasets Used}
\label{sec:datasets-used-virial-analysis}

\subsubsection{Dust Continuum-based Properties - Mass and Size}
\label{sec:datasets-core-characteristics}

We adopt the dataset from \citet{Belloche2011} for the Chamaeleon I cloud, \citet{Jorgensen2008} for Ophiuchus, and \citetalias{Nutter2007} for Orion B North, which are sub-millimetre dust continuum observations from single dish telescopes deriving the mass and size of the dense core population.

Chamaeleon I observations were conducted with the LABOCA bolometer array on the APEX telescope, measuring the dust continuum at 870~$\mu$m \citep{Belloche2011}. Cores were identified using \texttt{Gaussclumps} \citep{Stutzki1990, Kramer1998}, on the continuum maps of Chamaeleon I produced by \citet{Motte1998, Motte2007}.

Both the Ophiuchus and Orion B North datasets were observed with the SCUBA instrument on the JCMT at a wavelength of 850~$\mu$m \citep{Nutter2007, Jorgensen2008}. In the case of Ophiuchus, the core finding algorithm \texttt{clumpfind} was utilized \citep{Williams1994} and the core sizes were estimated based on the number of pixels contained within the \texttt{clumpfind} core boundary. For Orion B North, \citetalias{Nutter2007} estimated the core sizes, in most cases, by placing an elliptical aperture on each source and approximately matched to the position of the 3$\sigma$ contour. We directly adopt the masses and sizes as reported in these aforementioned papers for our analysis.

\subsubsection{Dense Gas Kinematics}
\label{sec:datasets-kinematic}

NH$_3$ and similar other N-bearing species are often found to be a good tracer of dense core material and can usually probe deeper layers than similar measurements conducted with C-bearing species \citep[e.g.,][]{Jijina1999, DiFrancesco2007}. These measurements provide similar measures of kinematic core properties across different star forming environments \citep{Johnstone2010}. The Green Bank Ammonia Survey \citep[GAS,][]{Friesen2017} mapped NH$_3(1,1)$ and NH$_3(2,2)$ in the Orion B North and Ophiuchus clouds, providing measurements of the dense gas excitation temperature and thermal and non-thermal components of the line width.

\citet{Tsitali2015} observed many different spectral lines in the Chamaeleon I cores; we adopt the N$_{2}$H$^{+}(1-0)$ data as being the most comparable to the NH$_3$ data available for the other two clouds. In the study of dense gas tracers in Perseus, \citet{Johnstone2010} show that the kinematic properties of NH$_3(1,1)$ and N$_{2}$H$^{+}(1,0)$ are extremely similar, showing that these species are well coupled, regardless of the physical conditions in the dense core gas. This reinforces our choice to only use the N$_{2}$H$^{+}(1-0)$ for the Chamaeleon I region. We assume a dense core gas temperature of $T=10~\text{K}$ for the Chamaeleon I cores, the same as \citet{Tsitali2015}.

\subsubsection{Larger Scale Turbulent Material}
\label{sec:dataset-turbulence}

The turbulent motions present in the lower density material surrounding the dense cores is often traced by various CO isotopologues. The all-sky CO surveys\footnote{The CO Survey Archive is found at the following: \url{https://lweb.cfa.harvard.edu/rtdc/CO/}} using the CfA and Cerro Tololo telescopes provides the best source of comparable CO data across all three regions studied \citep{Dame2001, Dame2022}. We choose to utilize the newer Northern Sky Survey of \citet{Dame2022}, which is a large extension of the Galactic plane CO survey of \citet{Dame2001} to the entire northern sky ($\delta > -17^{\circ}$). This survey uniformly samples the high-latitude sky with a resolution of 0.25~degrees, along with spectra with uniform sensitivity of 0.18~K in 0.65~km~s$^{-1}$ channels across a velocity range of $\pm$47.1~km~s$^{-1}$. We expect these contributions to lead to an overestimate of the associated linewidth due to the coarse spectral resolution; we discuss these implications in Section \ref{sec:virial-analysis}.

\subsubsection{Cloud Weight}
\label{sec:datasets-extinction-maps}

Finally, we utilize the all-sky dust extinction maps of \citet{Rowles2009} to estimate the compression on the dense cores due to the weight of their surrounding molecular cloud. We use the $N=25$ star maps to maximize the angular resolution available \citep[see Section 2.1 in][]{Rowles2009}.

\subsection{Data Coverage}
\label{sec:data-coverage}

The chosen extinction and $^{12}$CO maps are all-sky, therefore have uniform coverage over all three clouds studied. The dense gas tracers, NH$_3$ and N$_{2}$H$^{+}$, are more limited in terms of area mapped and area detected. We briefly explain the kinematic dataset coverage with respect to each dense core population studied.

\subsubsection{Ophiuchus}
\label{sec:data-coverage-oph}

GAS provides measurements of the dense gas line width and temperature, however, the latter quantity covers a smaller area as the fainter NH$_3(2,2)$ line needs to be detected in addition to the NH$_3(1,1)$ transition. In order to maximize the kinematic data available, we assume a temperature of $T=13~\text{K}$ for all cores that have only line width information available (taken as the mean temperature for all dense cores covered by GAS, in Ophiuchus). In Ophiuchus, 55 of the 66 dense cores, or 83\% have GAS line width measurements, with a subset of 29 cores containing kinetic temperature measurements. This dataset represents the largest overlap with the GAS kinematic data in our three cloud comparison.

Virial analyses of Ophiuchus based on SCUBA-2 data, have recently been published \citep{Pattle2015, Kerr2019}, however, for consistency with the SCUBA-based core catalogue used for the ALMA survey, we run our own independent analysis.

\subsubsection{Orion B North}
\label{sec:data-coverage-ori}

In Orion B North, only 18 of the 58 dense cores, or 31\% have GAS line width measurements, with a subset of 16 cores containing kinetic temperature measurements. The GAS mapping of Orion B North appears incomplete, resulting in a low amount of coverage in comparison to Ophiuchus. We note that the 16 cores with kinematic information available for this analysis are representative of the overall dense core population in Orion B North. We assume a temperature of $T=15~\text{K}$ for all cores which only have line width measurements (again taken as the mean temperature for all dense cores covered by GAS, in Orion B North).

\subsubsection{Chamaeleon I}
\label{sec:data-coverage-cha1}

In Chamaeleon I, 19 of the 60 dense cores, or 33\%, are detected in the N$_{2}$H$^{+}(1-0)$ line \citep{Tsitali2015}. \citet{Tsitali2015} do detect many more of the cores in lower density gas tracers, however, as noted in Section \ref{sec:datasets-kinematic}, we adopt only the N$_{2}$H$^{+}(1-0)$ data, as being the most comparable to the data available in the other two regions.

In total, we present the 19 dense cores in Chamaeleon I, 55 dense cores in Ophiuchus, and 18 dense cores in Orion B North, for a total of 92 dense cores in our virial analysis. We utilize the protostellar re-classification conducted by \citet{Dunham2016} for Chamaeleon I, \citet{Kirk2017} for Ophiuchus, and this work for Orion B North (see Table \ref{tab:observation-noise-levels} and Section \ref{sec:associations-summary}) to separate the dense core population into starless cores and protostellar cores as needed for the analysis.

\subsection{Jeans Analysis}
\label{sec:jeans-analysis}

We first compute the virial parameter, $\alpha$, which indicates the relative contribution of self-gravity versus that of thermal and non-thermal support. It is computed for a uniform, ellipsoid source, following the standard method outlined in \citet{Bertoldi1992}:

\begin{equation}\label{eq:virial-parameter}
    \alpha = \frac{5 \sigma_{\text{tot}}^2 R}{GM} \ ,
\end{equation}

\noindent where $R$ is the core radius, $G$ is the gravitational constant, and $M$ is the core mass.  The total velocity dispersion, $\sigma_{\text{tot}}$, is given by the following:

\begin{equation}\label{eq:total-velocity-disperson}
    \sigma_{\text{tot}} = \sqrt{(\sigma_{\text{obs}}^2 - \sigma_{\text{th,mol}}^2 + \sigma_{\text{th,mean}}^2)} \ ,
\end{equation}

\noindent with $\sigma_{\text{obs}}$ representing the observed dense gas line width. The mean and molecular velocity dispersions are calculated as follows:

\begin{equation}\label{eq:mean-velocity-dispersion}
    \sigma_{\text{th,mean}} = \sqrt{\frac{k_B T}{\mu m_H}} \ ,
\end{equation}

\begin{equation}\label{eq:molecular-velocity-dispersion}
    \sigma_{\text{th,mol}} = \sqrt{\frac{k_B T}{m_{\text{mol}}}} \ ,
\end{equation}

\noindent where the molecular weight of the mean particle is $\mu m_H$ ($\mu =2.37$), and $m_{\text{mol}} = \mu_{\text{mol}}m_H$ is the molecular mass of the tracer, in this case NH$_3$. As highlighted in Section \ref{sec:datasets-kinematic}, we compute these values for the Orion B North and Ophiuchus regions from the GAS data. We use the fitted line width and temperature values for the pixel that the core's peak lies within, where available. We utilize the published values by \citet{Tsitali2015} for their pointed observations of the Chamaeleon I cores from \citet{Belloche2011} and assume the same temperature of $T=10~\text{K}$, as outlined in Section \ref{sec:datasets-kinematic}.

\begin{figure*}
    \plotone{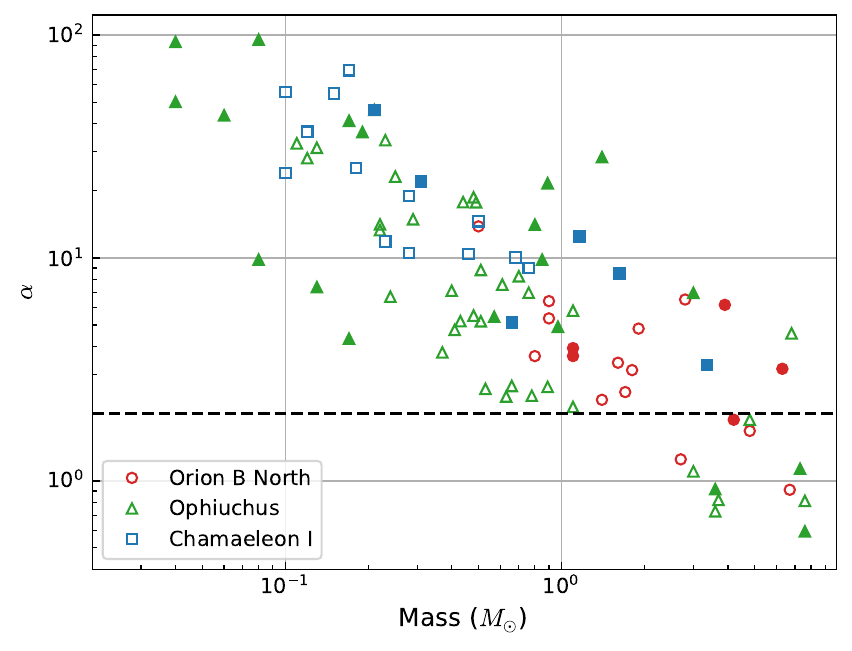}
    \caption{
    Virial parameter versus core mass, computed for each of the three cloud dense core populations. Each color represents a distinct star forming region. Open markers show the \emph{starless} core population, while filled markers show cores that are protostellar in nature. Cores lying above the dashed line are considered gravitationally unbound.
    \label{fig:jeans-analysis}
    }
\end{figure*}

Figure \ref{fig:jeans-analysis} plots the virial parameter against the core mass for all three core populations. None of the Chamaeleon I dense cores fall below the dashed line, indicating that all cores are gravitationally unbounded under this simple virial analysis. In contrast, 15\% ($8/55$) and 22\% ($4/18$) dense cores have $\alpha \le 2$ in Ophiuchus and Orion B North respectively. Many of the aforementioned cores that are found to be bounded are also starless in nature. The low ratio of Class $0+$I to Class II YSOs in Chamaeleon I, as presented in \citet{Dunham2015}, indicates a decelerating rate of star formation as compared to many other nearby star-forming regions, and along with a lack of bound cores, could explain the lack of detections of starless core substructure seen in \citet{Dunham2016}.

In Ophiuchus, \citet{Kerr2019} found that the majority of dense cores were found to be unbounded when only considering self-gravity and thermal and non-thermal motions. Only 7\% of such cores ($5/74$) had $\alpha \le 2$ in the Ophiuchus region \citep{Kerr2019}, presenting a slightly lower percentage of cores compared to this study's 15\%. Due to the different dust-based core catalogs used for the analysis, identical results are not expected. 

In summary, due to the characteristically different results between the Chamaeleon I cloud and the others clouds in this study, we conduct a broader look at further physical mechanisms through additional virial terms.

\subsection{Virial Analysis}
\label{sec:virial-analysis}

A number of recent studies implement a more complete virial analysis, often case in terms of energy densities, to gain a fuller picture of the core boundedness \citep{Pattle2015, Keown2017, Kirk2017a, Kerr2019}. Similar to previous studies, we incorporate both the external pressure contributions from the weight of the surrounding molecular cloud, as well as the pressure from the surrounding turbulent medium, but lack uniform data for the contributions of magnetic fields (as introduced in Section \ref{sec:comparisons}).

Following the method outlined in \citet{Pattle2015}, we describe the contributions given by internal motions and self-gravity in the core, as well as the external pressure contributions from the surrounding cloud as the following:

\begin{equation}\label{eq:virial_k}
    \Omega_{\text{k}} = \frac{3}{2}M\sigma_{\text{tot}}^2 \ ,
\end{equation}

\begin{equation}\label{eq:virial_g}
    \Omega_{\text{g}} = -\frac{3}{5} \frac{GM^2}{R} \ ,
\end{equation}

\begin{equation}\label{eq:virial_p}
    \Omega_{\text{P}} = -4\pi PR^3 \ .
\end{equation}

We follow \citet{Kerr2019} and choose a constant factor of $-3/5$ for Equation \ref{eq:virial_g}, appropriate for sources with constant density, which is a better approximation for cores following the typical $\rho \propto r^{-1}$ density profile. 

The contribution of the molecular cloud weight is computed with the following \citep{McKee1989, Kirk2017a}:

\begin{equation}\label{eq:cloud-weight-pressure}
    P_{\text{w}} = \pi G \Sigma \bar{\Sigma} \ ,
\end{equation}

\noindent where $\Sigma$ is the surface mass density along the line-of-sight of the dense core, and $\bar{\Sigma}$ is the mean mass surface density across the entire cloud. This formulation assumes that the large scale structure is spherically symmetric, with a density falloff following $\rho \propto r^{-1}$; the effect of different geometries is discussed in \citet{Kirk2017a}. We compute the mean mass surface density using all pixels above $A_{\text{v}} = 3~\text{mag}$ associated with the cloud. We convert the extinction into column density using a conversion of $1~\text{mag} = 9.4\times 10^{20}~\text{cm}^{-2}$ from \citet{Bohlin1978}. The mean extinctions are similar across all three clouds, with values of 4.95~mag, 5.23~mag, and 4.91~mag, for Chamaeleon I, Ophiuchus, and Orion B North respectively. Due to the use of comparatively lower resolution data, the contribution of core material to the extinction measurement is minimal, and further processing of the extinction maps are not required.

To incorporate the contribution from large-scale turbulent pressure, we calculate the line widths of the all-sky CO data (see Section \ref{sec:dataset-turbulence}) at the position of each core. We utilize a single component Gaussian model, which is a reasonable representation based on visual checks of the spectra. Assuming the turbulent pressure to be isotropic in nature, we use the following expression:

\begin{equation}\label{eq:turbulence-pressure}
    P_{\text{t}} = n_{\text{CO}} \mu_{\text{mn}} m_{\text{H}} \sigma_{\text{tot,CO}}^2 \ ,
\end{equation}

\noindent where the density of the gas that the $^{12}$CO lines are probing is given by $n_{\text{CO}}$, and the fitted linewidth is given by $\sigma_{\text{tot,CO}}$. Due to the lower angular resolution of the $^{12}$CO data (0.25~degrees), we utilize the mean density found in star forming cloud environments, 250~cm$^{-3}$ \citep{Bergin2007} to represent the density of the gas that the $^{12}$CO data is probing.

\startlongtable
\begin{deluxetable*}{cccccccccl}
    \tablecolumns{10}
    \tablecaption{Virial Analysis - Starless Core Properties \label{tab:virial-analysis}}
    \tablehead{
        \colhead{Core Name\tablenotemark{a}} & 
        \colhead{Mass\tablenotemark{a}} & 
        \colhead{$R_{\text{eff}}$\tablenotemark{a}} & 
        \colhead{$T_k$\tablenotemark{b}} & 
        \colhead{$\sigma_{\text{obs}}$\tablenotemark{b}} & 
        \colhead{$\Omega_{\text{k}}$\tablenotemark{c}} & 
        \colhead{$-\Omega_{\text{g}}$\tablenotemark{c}} & 
        \colhead{$-\Omega_{\text{P,w}}$\tablenotemark{c}} & 
        \colhead{$-\Omega_{\text{P,t}}$\tablenotemark{c}} & 
        \colhead{Virial\tablenotemark{d}} \\
        \nocolhead{} & 
        \colhead{($M_{\odot}$)} & 
        \colhead{(pc)} & 
        \colhead{(K)} & 
        \colhead{($\text{km}~\text{s}^{-1}$)} & 
        \colhead{(erg)} & 
        \colhead{(erg)} & 
        \colhead{(erg)} & 
        \colhead{(erg)} & 
        \colhead{Ratio}
    }
    \startdata
    \sidehead{Chamaeleon I}
    Cha1-C1 & 3.36 & 0.035 & 10.0 & 0.490 & 2.71e+43 & 1.64e+43 & 4.37e+41 & 1.33e+42 & 3.35e-01 \\
    Cha1-C2 & 1.62 & 0.046 & 10.0 & 0.480 & 1.26e+43 & 2.95e+42 & 8.79e+41 & 2.86e+42 & 2.66e-01 \\
    Cha1-C3 & 1.16 & 0.040 & 10.0 & 0.530 & 1.08e+43 & 1.72e+42 & 6.82e+41 & 1.94e+42 & 2.02e-01 \\
    Cha1-C4 & 0.66 & 0.035 & 10.0 & 0.230 & 1.63e+42 & 6.35e+41 & 4.51e+41 & 1.55e+42 & 8.08e-01 \\
    Cha1-C8 & 0.15 & 0.019 & 10.0 & 0.580 & 1.64e+42 & 5.99e+40 & 7.57e+40 & 2.16e+41 & 1.07e-01 \\
    Cha1-C9 & 0.68 & 0.044 & 10.0 & 0.320 & 2.68e+42 & 5.35e+41 & 4.47e+41 & 2.28e+42 & 6.07e-01 \\
    Cha1-C10 & 0.31 & 0.032 & 10.0 & 0.390 & 1.68e+42 & 1.53e+41 & 3.17e+41 & 1.01e+42 & 4.40e-01 \\
    Cha1-C11 & 0.76 & 0.052 & 10.0 & 0.290 & 2.59e+42 & 5.73e+41 & 1.34e+42 & 4.69e+42 & 1.28e+00 \\
    Cha1-C14 & 0.21 & 0.028 & 10.0 & 0.520 & 1.88e+42 & 8.15e+40 & 2.37e+41 & 6.45e+41 & 2.56e-01 \\
    Cha1-C19 & 0.28 & 0.036 & 10.0 & 0.310 & 1.05e+42 & 1.11e+41 & 5.64e+41 & 1.49e+42 & 1.03e+00 \\
    Cha1-C21 & 0.10 & 0.022 & 10.0 & 0.250 & 2.76e+41 & 2.30e+40 & 1.19e+41 & 3.46e+41 & 8.84e-01 \\
    Cha1-C29 & 0.17 & 0.029 & 10.0 & 0.570 & 1.80e+42 & 5.18e+40 & 2.15e+41 & 7.08e+41 & 2.71e-01 \\
    Cha1-C30 & 0.46 & 0.050 & 10.0 & 0.230 & 1.14e+42 & 2.18e+41 & 9.21e+41 & 4.19e+42 & 2.34e+00 \\
    Cha1-C31 & 0.28 & 0.038 & 10.0 & 0.190 & 5.52e+41 & 1.05e+41 & 4.87e+41 & 1.91e+42 & 2.27e+00 \\
    Cha1-C33 & 0.18 & 0.031 & 10.0 & 0.310 & 6.77e+41 & 5.35e+40 & 3.02e+41 & 1.02e+42 & 1.01e+00 \\
    Cha1-C34 & 0.50 & 0.052 & 10.0 & 0.300 & 1.79e+42 & 2.45e+41 & 8.44e+41 & 3.30e+42 & 1.23e+00 \\
    Cha1-C35 & 0.12 & 0.026 & 10.0 & 0.340 & 5.21e+41 & 2.83e+40 & 2.93e+41 & 7.07e+41 & 9.87e-01 \\
    Cha1-C40 & 0.23 & 0.035 & 10.0 & 0.190 & 4.53e+41 & 7.65e+40 & 5.30e+41 & 1.59e+42 & 2.42e+00 \\
    Cha1-C42 & 0.10 & 0.024 & 10.0 & 0.410 & 5.91e+41 & 2.12e+40 & 1.55e+41 & 4.25e+41 & 5.09e-01 \\
    \sidehead{Ophiuchus}
    162608-24202 & 0.76 & 0.028 & 13.0 & 0.355 & 3.74e+42 & 1.07e+42 & 5.25e+41 & 5.42e+41 & 2.86e-01 \\
    162610-24195 & 0.48 & 0.023 & 13.0 & 0.551 & 4.91e+42 & 5.24e+41 & 2.78e+41 & 2.94e+41 & 1.12e-01 \\
    162610-24231 & 0.44 & 0.023 & 13.0 & 0.509 & 3.92e+42 & 4.41e+41 & 2.85e+41 & 2.94e+41 & 1.30e-01 \\
    162610-24206 & 0.80 & 0.032 & 13.0 & 0.514 & 7.24e+42 & 1.03e+42 & 8.34e+41 & 8.42e+41 & 1.87e-01 \\
    162614-24232 & 0.13 & 0.012 & 13.0 & 0.493 & 1.09e+42 & 7.01e+40 & 5.35e+40 & 4.85e+40 & 7.87e-02 \\
    162614-24234 & 0.25 & 0.018 & 13.0 & 0.485 & 2.04e+42 & 1.76e+41 & 1.58e+41 & 1.54e+41 & 1.20e-01 \\
    162614-24250 & 1.40 & 0.039 & 13.0 & 0.920 & 3.69e+43 & 2.61e+42 & 1.69e+42 & 1.47e+42 & 7.80e-02 \\
    162615-24231 & 0.49 & 0.025 & 13.0 & 0.514 & 4.43e+42 & 4.98e+41 & 4.28e+41 & 3.88e+41 & 1.49e-01 \\
    162616-24235 & 0.23 & 0.017 & 13.0 & 0.586 & 2.62e+42 & 1.56e+41 & 1.62e+41 & 1.37e+41 & 8.66e-02 \\
    162617-24235 & 0.89 & 0.031 & 13.0 & 0.709 & 1.44e+43 & 1.33e+42 & 8.92e+41 & 7.32e+41 & 1.03e-01 \\
    162622-24225 & 3.00 & 0.042 & 13.0 & 0.622 & 3.81e+43 & 1.09e+43 & 1.98e+42 & 1.93e+42 & 1.95e-01 \\
    162624-24162 & 0.06 & 0.007 & 13.0 & 0.520 & 5.54e+41 & 2.54e+40 & 8.21e+39 & 1.30e+40 & 4.21e-02 \\
    162626-24243 & 7.30 & 0.048 & 18.3 & 0.305 & 3.22e+43 & 5.70e+43 & 3.44e+42 & 2.84e+42 & 9.81e-01 \\
    162627-24233 & 3.00 & 0.023 & 18.3 & 0.259 & 1.09e+43 & 1.98e+43 & 3.34e+41 & 3.24e+41 & 9.38e-01 \\
    162628-24235 & 7.60 & 0.043 & 18.5 & 0.261 & 2.80e+43 & 6.91e+43 & 2.46e+42 & 2.03e+42 & 1.31e+00 \\
    162628-24225 & 4.80 & 0.051 & 18.1 & 0.313 & 2.18e+43 & 2.32e+43 & 3.50e+42 & 3.39e+42 & 6.90e-01 \\
    162633-24261 & 6.80 & 0.073 & 13.0 & 0.574 & 7.48e+43 & 3.26e+43 & 1.23e+43 & 9.65e+42 & 3.64e-01 \\
    162641-24272 & 0.85 & 0.033 & 13.0 & 0.425 & 5.56e+42 & 1.13e+42 & 1.21e+42 & 8.79e+41 & 2.90e-01 \\
    162644-24173 & 0.22 & 0.015 & 13.0 & 0.367 & 1.14e+42 & 1.71e+41 & 7.99e+40 & 1.04e+41 & 1.56e-01 \\
    162644-24345 & 0.04 & 0.007 & 13.0 & 0.445 & 2.83e+41 & 1.13e+40 & 9.72e+39 & 9.65e+39 & 5.42e-02 \\
    162644-24253 & 0.41 & 0.023 & 14.8 & 0.173 & 9.09e+41 & 3.83e+41 & 3.69e+41 & 2.87e+41 & 5.71e-01 \\
    162646-24242 & 0.40 & 0.022 & 13.0 & 0.271 & 1.34e+42 & 3.76e+41 & 3.44e+41 & 2.60e+41 & 3.66e-01 \\
    162648-24236 & 0.51 & 0.026 & 13.0 & 0.330 & 2.25e+42 & 5.10e+41 & 5.95e+41 & 9.04e+41 & 4.47e-01 \\
    162660-24343 & 3.60 & 0.053 & 10.3 & 0.109 & 4.57e+42 & 1.25e+43 & 4.30e+42 & 3.75e+42 & 2.25e+00 \\
    162705-24363 & 0.08 & 0.010 & 13.0 & 0.779 & 1.54e+42 & 3.23e+40 & 2.60e+40 & 2.65e+40 & 2.75e-02 \\
    162705-24391 & 0.53 & 0.023 & 11.4 & 0.128 & 8.00e+41 & 6.19e+41 & 2.79e+41 & 3.16e+41 & 7.59e-01 \\
    162707-24381 & 0.04 & 0.007 & 13.0 & 0.634 & 5.26e+41 & 1.13e+40 & 8.51e+39 & 9.65e+39 & 2.80e-02 \\
    162709-24372 & 0.19 & 0.015 & 13.0 & 0.611 & 2.34e+42 & 1.27e+41 & 6.80e+40 & 7.72e+40 & 5.83e-02 \\
    162711-24393 & 0.21 & 0.017 & 13.0 & 0.678 & 3.12e+42 & 1.35e+41 & 1.15e+41 & 1.18e+41 & 5.90e-02 \\
    162712-24290 & 0.11 & 0.012 & 15.2 & 0.469 & 8.72e+41 & 5.34e+40 & 5.13e+40 & 3.95e+40 & 8.27e-02 \\
    162712-24380 & 0.12 & 0.013 & 13.0 & 0.427 & 7.91e+41 & 5.64e+40 & 5.68e+40 & 5.63e+40 & 1.07e-01 \\
    162713-24295 & 0.29 & 0.017 & 15.6 & 0.408 & 1.84e+42 & 2.47e+41 & 1.61e+41 & 1.33e+41 & 1.47e-01 \\
    162715-24303 & 0.22 & 0.017 & 15.3 & 0.338 & 1.05e+42 & 1.48e+41 & 1.27e+41 & 1.17e+41 & 1.87e-01 \\
    162725-24273 & 0.70 & 0.023 & 13.4 & 0.425 & 4.61e+42 & 1.12e+42 & 3.03e+41 & 5.77e+41 & 2.16e-01 \\
    162727-24405 & 3.60 & 0.055 & 12.2 & 0.121 & 5.51e+42 & 1.20e+43 & 3.67e+42 & 4.24e+42 & 1.81e+00 \\
    162728-24271 & 0.97 & 0.022 & 13.3 & 0.384 & 5.43e+42 & 2.21e+42 & 2.74e+41 & 5.23e+41 & 2.77e-01 \\
    162728-24393 & 0.17 & 0.015 & 13.0 & 0.613 & 2.10e+42 & 1.02e+41 & 7.58e+40 & 7.73e+40 & 6.07e-02 \\
    162729-24274 & 0.57 & 0.023 & 13.0 & 0.282 & 2.01e+42 & 7.40e+41 & 2.83e+41 & 5.77e+41 & 3.97e-01 \\
    162730-24264 & 0.51 & 0.019 & 13.7 & 0.282 & 1.83e+42 & 7.06e+41 & 1.70e+41 & 3.41e+41 & 3.32e-01 \\
    162730-24415 & 0.61 & 0.029 & 13.0 & 0.313 & 2.49e+42 & 6.56e+41 & 5.58e+41 & 6.19e+41 & 3.68e-01 \\
    162733-24262 & 1.10 & 0.028 & 14.1 & 0.395 & 6.51e+42 & 2.25e+42 & 5.20e+41 & 1.06e+42 & 2.94e-01 \\
    162739-24424 & 0.37 & 0.023 & 11.8 & 0.133 & 5.87e+41 & 3.12e+41 & 2.61e+41 & 2.88e+41 & 7.34e-01 \\
    162740-24431 & 0.13 & 0.013 & 13.0 & 0.156 & 2.45e+41 & 6.62e+40 & 5.12e+40 & 5.64e+40 & 3.55e-01 \\
    162759-24334 & 0.63 & 0.025 & 11.0 & 0.134 & 9.55e+41 & 8.00e+41 & 3.57e+41 & 6.90e+41 & 9.68e-01 \\
    162821-24362 & 0.17 & 0.014 & 11.5 & 0.108 & 2.33e+41 & 1.07e+41 & 5.33e+40 & 1.10e+41 & 5.82e-01 \\
    163138-24495 & 1.10 & 0.031 & 12.2 & 0.173 & 2.18e+42 & 2.03e+42 & 3.79e+41 & 5.74e+41 & 6.85e-01 \\
    163139-24506 & 0.48 & 0.026 & 12.7 & 0.221 & 1.24e+42 & 4.52e+41 & 2.53e+41 & 3.62e+41 & 4.29e-01 \\
    163140-24485 & 0.24 & 0.018 & 9.9 & 0.215 & 5.45e+41 & 1.63e+41 & 8.09e+40 & 1.21e+41 & 3.35e-01 \\
    163141-24495 & 0.89 & 0.029 & 12.2 & 0.182 & 1.84e+42 & 1.40e+42 & 3.79e+41 & 4.96e+41 & 6.16e-01 \\
    163154-24560 & 0.43 & 0.023 & 11.7 & 0.218 & 1.06e+42 & 4.08e+41 & 1.87e+41 & 2.54e+41 & 4.00e-01 \\
    163157-24572 & 0.78 & 0.031 & 11.1 & 0.136 & 1.20e+42 & 9.98e+41 & 3.87e+41 & 6.63e+41 & 8.53e-01 \\
    163201-24564 & 0.08 & 0.009 & 13.9 & 0.189 & 1.85e+41 & 3.76e+40 & 9.39e+39 & 1.44e+40 & 1.66e-01 \\
    163223-24284 & 7.60 & 0.041 & 13.8 & 0.231 & 2.15e+43 & 7.28e+43 & 1.44e+42 & 1.68e+42 & 1.76e+00 \\
    163229-24291 & 3.70 & 0.047 & 12.1 & 0.141 & 6.19e+42 & 1.51e+43 & 2.22e+42 & 2.50e+42 & 1.60e+00 \\
    163448-24381 & 0.66 & 0.031 & 8.8 & 0.152 & 9.74e+41 & 7.32e+41 & 3.45e+41 & 3.06e+41 & 7.10e-01 \\
    \sidehead{Orion B North}
    BN-546074-01342 & 6.30 & 0.068 & 17.1 & 0.449 & 4.75e+43 & 2.99e+43 & 3.57e+42 & 1.45e+43 & 5.04e-01 \\
    BN-546097-00552 & 0.50 & 0.035 & 15.0 & 0.353 & 2.53e+42 & 3.65e+41 & 5.42e+41 & 1.98e+42 & 5.70e-01 \\
    BN-546135-00525 & 1.10 & 0.045 & 12.8 & 0.213 & 2.75e+42 & 1.40e+42 & 1.32e+42 & 4.02e+42 & 1.23e+00 \\
    BN-546310-00234 & 4.20 & 0.053 & 16.1 & 0.284 & 1.61e+43 & 1.72e+43 & 2.34e+42 & 6.66e+42 & 8.11e-01 \\
    BN-546433-00148 & 0.80 & 0.049 & 12.3 & 0.119 & 1.22e+42 & 6.72e+41 & 1.31e+42 & 5.61e+42 & 3.11e+00 \\
    BN-546280-00145 & 1.90 & 0.043 & 13.7 & 0.379 & 1.05e+43 & 4.34e+42 & 1.44e+42 & 3.54e+42 & 4.45e-01 \\
    BN-546276-00057 & 3.90 & 0.047 & 15.5 & 0.630 & 5.16e+43 & 1.68e+43 & 1.96e+42 & 4.60e+42 & 2.26e-01 \\
    BN-546321-00044 & 0.90 & 0.035 & 12.8 & 0.320 & 3.78e+42 & 1.18e+42 & 8.10e+41 & 1.99e+42 & 5.26e-01 \\
    BN-546532-00018 & 4.80 & 0.070 & 16.7 & 0.221 & 1.41e+43 & 1.69e+43 & 3.42e+42 & 1.65e+43 & 1.30e+00 \\
    BN-546334-00006 & 1.10 & 0.038 & 15.4 & 0.210 & 2.95e+42 & 1.63e+42 & 8.09e+41 & 2.52e+42 & 8.40e-01 \\
    BN-546244-00001 & 2.80 & 0.059 & 15.0 & 0.470 & 2.22e+43 & 6.83e+42 & 3.87e+42 & 1.11e+43 & 4.90e-01 \\
    BN-546450+00021 & 1.70 & 0.036 & 16.0 & 0.234 & 5.22e+42 & 4.17e+42 & 5.08e+41 & 2.16e+42 & 6.56e-01 \\
    BN-546350+00029 & 1.60 & 0.044 & 17.0 & 0.235 & 5.06e+42 & 2.99e+42 & 1.17e+42 & 4.61e+42 & 8.66e-01 \\
    BN-546405+00032 & 2.70 & 0.042 & 15.4 & 0.153 & 5.60e+42 & 8.99e+42 & 8.17e+41 & 3.47e+42 & 1.18e+00 \\
    BN-546267+00101 & 1.80 & 0.045 & 13.6 & 0.259 & 5.77e+42 & 3.68e+42 & 1.32e+42 & 5.00e+42 & 8.65e-01 \\
    BN-546275+00135 & 1.40 & 0.038 & 13.1 & 0.183 & 3.04e+42 & 2.64e+42 & 7.93e+41 & 3.00e+42 & 1.06e+00 \\
    BN-546499+00204 & 6.70 & 0.068 & 17.7 & 0.155 & 1.54e+43 & 3.38e+43 & 3.05e+42 & 1.53e+43 & 1.69e+00 \\
    BN-546362+00550 & 0.90 & 0.035 & 15.0 & 0.270 & 3.16e+42 & 1.18e+42 & 7.36e+41 & 2.87e+42 & 7.57e-01 \\
    \enddata
    \tablenotetext{a}{Dense core name, mass, and effective radius as adopted from \citet{Belloche2011, Jorgensen2008, Nutter2007} for Chamaeleon I, Ophiuchus, and Orion B North respectively.}
    \tablenotetext{b}{Observed linewidth and kinetic temperature as measured from the GAS kinematic dataset. We assume a temperature of $T=13~\text{K}$ and $T=15~\text{K}$ for Ophiuchus and Orion B North respectively, for all cores which only have linewidth measurements (see Section \ref{sec:data-coverage} for discussions).}
    \tablenotetext{c}{Virial parameters as derived in Section \ref{sec:virial-analysis}.}
    \tablenotetext{d}{Indicates whether the dense core is bound taking into account all known virial terms; $-(\Omega_{\text{g}} + \Omega_{\text{P,t}} + \Omega_{\text{P,w}})/{2\Omega_{\text{k}}}$.}
\end{deluxetable*}

\begin{figure*}
    \plotone{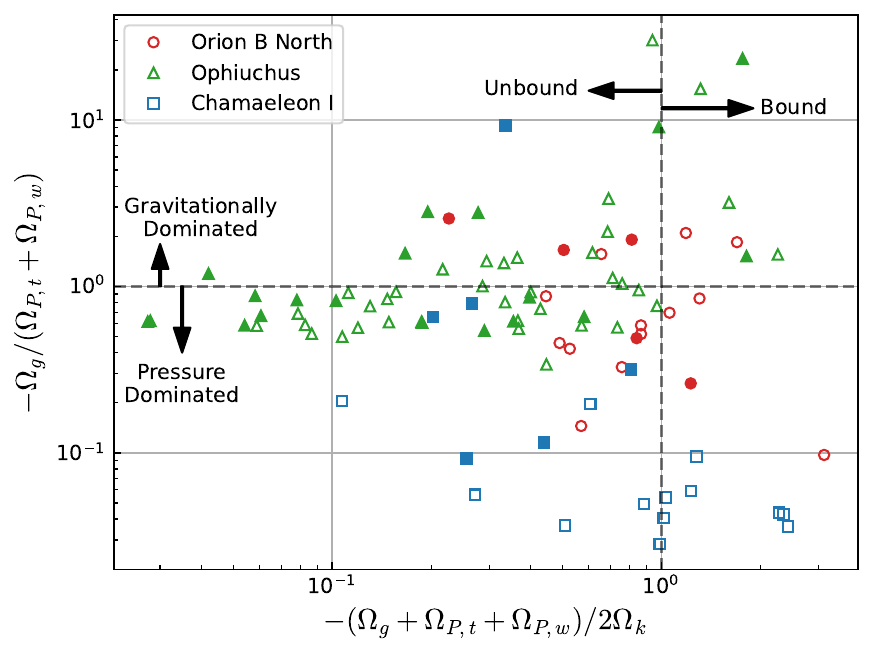}
    \caption{
    Confinement ratio versus the virial ratio, as computed for each of the three cloud dense core populations. See Figure \ref{fig:jeans-analysis} for plotting conventions. Cores lying above the horizontal dashed line are considered gravitationally dominated, while cores under the dashed line are considered pressure dominated. Cores lying to the right hand side of the vertical dashed line are bounded given the full virial equation used for this study.
    \label{fig:virial-analysis}
    }
\end{figure*}

Our virial results are presented in Table \ref{tab:virial-analysis}, for the total 92 dense cores across the three star forming regions. A dense core is defined as being in a state of virial equilibrium when $2\Omega_{\text{k}} = -(\Omega_{\text{g}} + \Omega_{\text{P,t}} + \Omega_{\text{P,w}})$. If the virial ratio (as given in the final column of Table \ref{tab:virial-analysis}) is greater than one, a core will be bounded, while a ratio less than one indicates that the core is unbounded and most likely will disperse over time. Figure \ref{fig:virial-analysis} shows the confinement ratio on the vertical axis, describing the relative contributions of both external pressure and self-gravity, as a function of the virial ratio.

The addition of the contributions of the external pressure terms, do generally shift more cores into the bound regime, as compared to our results from the Jeans analysis. The relative contribution of external pressure indicates that the overall cloud pressure is responsible for much more of the binding in Chamaeleon I cores, up to an order of magnitude on average compared with those found in the Ophiuchus region. Similar to \citet{Tsitali2015}, we find Cha1-C1 (also known as Cha-MMS1) to be gravitationally dominated, the only core in our analysis in this domain. Cha-MMS1 is a first hydrostatic core candidate \citep{Belloche2006, Belloche2011, Tsitali2013}, explaining the much larger contribution of the gravitational virial term in comparison its external pressure term. 

With the addition of turbulent and cloud weight pressures, 54\% ($7/13$) of starless cores in Chamaeleon I, 8\% ($3/36$) of starless cores in Ophiuchus, and 38\% ($5/13$) of cores in Orion B North qualify as bound. Compared to the results from \citet{Kerr2019}, in which 36\% of the cores were found to become bound under the use of all virial terms, we find a lower amount of cores in Ophiuchus that are classified as bound. We do show that most cores in Ophiuchus have more equal contributions from the two pressure sources, similar to the results from \citet{Kerr2019}.

The all-sky CO survey dataset from \citet{Dame2022} used in this analysis is chosen for its comparable observations and spectral resolution across the entire northern sky (see Section \ref{sec:dataset-turbulence}). The individual all-sky CO survey datasets for Chamealeon \citep{Boulanger1998}, Ophiuchus \citep{DeGeus1990}, and Orion \citep{Wilson2005} do give a higher angular resolution for the Orion and Chamaeleon regions (0.125~degrees), and a higher spectral resolution for Ophiuchus (0.26~km~s$^{-1}$). Utilizing these individual surveys in-place of our chosen CO dataset yielded differences in the fitted linewidths of approximately 20\% on average, and no more than 30\% in difference. These specific changes in the fitted spectra do not significantly change the results shown. In summary, due to the coarse spectral resolution of the surveys (as compared to other similar studies), we are likely to have over-estimated our turbulent pressures across all three core populations.

For typical Milky Way cloud conditions numerical simulations of magnetized, turbulent clouds suggest that $\sim$20-40\% of dense cores disperse before forming stars \citep{Smullen2020, Offner2022}.  Low-surface density, less massive clouds like Chamaeleon likely have higher fractions of dispersing starless cores. However, the measured virial parameter, as defined in Equation \ref{eq:virial-parameter}, poorly correlates with whether a given core will go on to collapse. \citet{Offner2022} analyze the relationship between a variety of core properties and their evolution and find that individual statistics, like virial parameter, are poor predictors of whether cores eventually become protostellar. This is perhaps not surprising as observational estimates of boundedness neglect important dynamical indicators, including time variation and the magnetic energy \citep{Dib2007}. However, a more holistic view of core properties, including their degree of velocity coherence, mass, size, and density/velocity profiles, enables more accurate predictions of core outcomes. By mapping core properties inferred from NH$_3$ observations to those of simulated cores, \citet{Offner2022} predict that at least $58 \pm 14\%$ and $55 \pm 12\%$ of Ophiuchus and Orion cores, respectively, go on to form stars. This suggests that predictions for detectable core substructure that posit 100\% of observed cores collapse, as assumed here, are reasonable approximations for these regions.

In summary, while the virial measures of each core may be a poor indicator of it's future state, the large difference in bulk population properties between Chamaeleon I and Ophiuchus/Orion B North may lend a plausible explanation for the lack of ALMA detections in the Chamaeleon I region.

\section{Conclusions}
\label{sec:conclusions}

In this paper, we present ALMA Cycle 3 Band 3 observations of 73 dense cores in the Orion B North region. We perform synthetic observations of starless core evolution simulations to predict the expected number of starless core detections in our dataset under the turbulent fragmentation model. We summarize our main results as follows:

\begin{enumerate}
    \item We detect 34 continuum sources across 19 individual ALMA pointings. Four of these detections are most likely starless, as we find no protostellar association nor signs of CO outflow at their location.
    \item The likely starless cores detected through our ALMA observations are among the faintest objects in our survey and all are associated with mid- to high-concentration parent cores, as measured by the lower-resolution \citetalias{Nutter2007} SCUBA data.
    \item We generate synthetic observations of isolated starless core evolution simulations to predict the number of starless core detections expected under the turbulent fragmentation model. Out of the 58 truly starless cores observed with ALMA, our one starless core detection agrees with the predicted number of detections $2\pm1$ under the turbulent fragmentation model.
\end{enumerate}

We put these results in the context of previous ALMA surveys by performing a multi-cloud virial analysis with dense core populations observed with ALMA. The Ophiuchus and Orion B North results are consistent with the turbulent fragmentation picture, while Chamaeleon I results are not. We perform an additional virial analysis on all three dense core populations, and find that Chamaeleon I shows features unlike both Ophiuchus and Orion B North. According to our simple Jeans analysis, no Chamaeleon I cores are bound. With the contributions of external pressure terms in our virial analysis, we show that for the Chamaeleon I cores, an order of magnitude more of the binding energy is attributed to cloud weight and cloud-scale turbulent pressure than self-gravity, compared with the other two clouds. These differences lend weight to the findings of \citet{Tsitali2015}, in which the dense cores in Chamaeleon I, unlike the majority of cores in most star-forming regions, are destined to re-expand without forming stars.

\section*{Acknowledgements}
We thank the referee for their insightful and quick reports which improved the manuscript. We acknowledge the helpful input from John Tobin, the external examiner for SF's MSc thesis, upon which this paper is based.
SF acknowledges and respects the Lekwungen-speaking peoples on whose traditional territories SF works and lives, and the Songhees, Esquimalt and WSANEC peoples whose historical relationships with the land continue to this day.
The National Radio Astronomy Observatory is a facility of the National Science Foundation operated under cooperative agreement by Associated Universities, Inc. This paper makes use of the following ALMA data, with project code: \#2015.1.00094.S. ALMA is a partnership of ESO (representing its member states), NSF (USA), and NINS (Japan), together with NRC (Canada) and NSC and ASIAA (Taiwan), in cooperation with the Republic of Chile. The Joint ALMA Observatory is operated by ESO, AUI/NRAO, and NAOJ. This research also made use of NASA’s Astrophysics Data System (ADS) Abstract Service. This research has made use of the SIMBAD database, operated at CDS, Strasbourg, France \citep{Wenger2000}. The JCMT has historically been operated by the Joint Astronomy Center on behalf of the Science and Technology Facilities Council of the United Kingdom, the National Research Council of Canada and the Netherlands Organisation for Scientific Research.
The authors acknowledge the use of the Canadian Advanced Network for Astronomy Research (CANFAR) Science Platform. Our work used the facilities of the Canadian Astronomy Data Center, operated by the National Research Council of Canada with the support of the Canadian Space Agency, and CANFAR, a consortium that serves the data-intensive storage, access, and processing needs of university groups and centers engaged in astronomy research \citep{Gaudet2010}. We thank the GAS Team for their publicly available DR1 dataset. SF and HK acknowledge support from an NSERC Discovery Grant. SO acknowledges support from NSF Career grant 1748571 and a Moncrief Grant Challenge Award.

\section*{Data Availability}
All associated data, including compiled figures and tables, are publicly available at the CADC via the following link: \url{http://doi.org/10.11570/24.0007} \citep{Fielder-CADC-DOI}.

\vspace{5mm}
\facilities{ALMA, JCMT, GBT}

\software{
\citep[Astropy][]{AstropyCollaboration2013, AstropyCollaboration2018, AstropyCollaboration2022}, \citep[CASA][]{TheCASATeam2022}
}

\appendix

\section{Comparison of SCUBA and SCUBA-2 data}
\label{sec:scuba-comparison}

While not available at the time of the ALMA proposal, the Orion molecular cloud was observed by the JCMT Gould Belt Survey with the SCUBA-2 instrument \citep{Kirk2016}, which has larger spatial coverage and improved sensitivity compared to the \citetalias{Nutter2007} SCUBA observations. The SCUBA \citetalias{Nutter2007} has a sensitivity of 16~mJy~beam$^{-1}$ at 850~$\mu$m, while comparatively, the SCUBA-2 observations by \citet{Kirk2016} has a sensitivity of 3.4~mJy~beam$^{-1}$ at the same wavelength. Additionally, two different core finding algorithms were used to identify over-dense structures in the emission maps, leading to slightly different core positions and footprints when looking at the same star forming region. For example, in Figure \ref{fig:Field0032}, the two ALMA observations for this paper (derived from the \citetalias{Nutter2007} catalog) lie approximately in between 3 distinct SCUBA-2 cores, as shown by the light blue contours over-plotted. While we use the SCUBA observations for most of our analysis, the SCUBA-2 cores are mentioned when this context is needed.

According to \citet{Kirk2016}, each peak flux position from the SCUBA dataset was associated with the SCUBA-2 core whose boundary it lies within, resulting in all 100 SCUBA cores finding an association with a SCUBA-2 core. Associated core values agree reasonably well between the two datasets, with some scatter present when looking at cores which have one-to-one associations \citep{Kirk2016}. Due to the increased sensitivity of the SCUBA-2 map and the different core-finding algorithm adopted, there is not a one-to-one correspondence between \textit{all} SCUBA and SCUBA-2 cores. In some cases, multiple SCUBA cores map to the same SCUBA-2 core, and in others, a single SCUBA core is subdivided into multiple SCUBA-2 cores.

By comparing both core catalogs, we find six instances where either two or three starless SCUBA cores correspond to a single SCUBA-2 core. In these instances, the multiple peaks identified in the SCUBA map lie along a shared filamentary structure, and the brightest of the SCUBA cores peaks in a similar location to the corresponding SCUBA-2 core peak. Meanwhile, the remaining SCUBA cores lie within the larger elongated footprint of the SCUBA-2 core rather than being associated with unique SCUBA-2 cores. These eight SCUBA cores that do not have unique SCUBA-2 core correspondences are potentially artificially subdivided components of a single larger structure rather than distinct cores. Additionally, we find three instances where individual SCUBA cores were found to lie inside of the associated SCUBA-2 core footprint but with separations between the SCUBA and SCUBA-2 peaks of 50-100 arcsec; we consider these tentatively non-correspondent. Finally, we find one instance where a SCUBA core has no corresponding SCUBA-2 core.

We additionally find three instances where individual SCUBA cores are each subdivided into two SCUBA-2 cores.

Using the SCUBA versus SCUBA-2 comparisons gives a guide to the accuracy and reliability in the cores identified, including their total number. We expect that the true number of distinct cores surveyed lies between the two extremes of the SCUBA and SCUBA-2 core comparisons described here. In summary, there may be as few as 47 or as high as 61 starless cores observed with ALMA.

\section{Synthetic Observations of Simulated $4M_{\odot}$ Starless Core Evolution}
\label{sec:4M_simulations}

\begin{figure*}
    \plotone{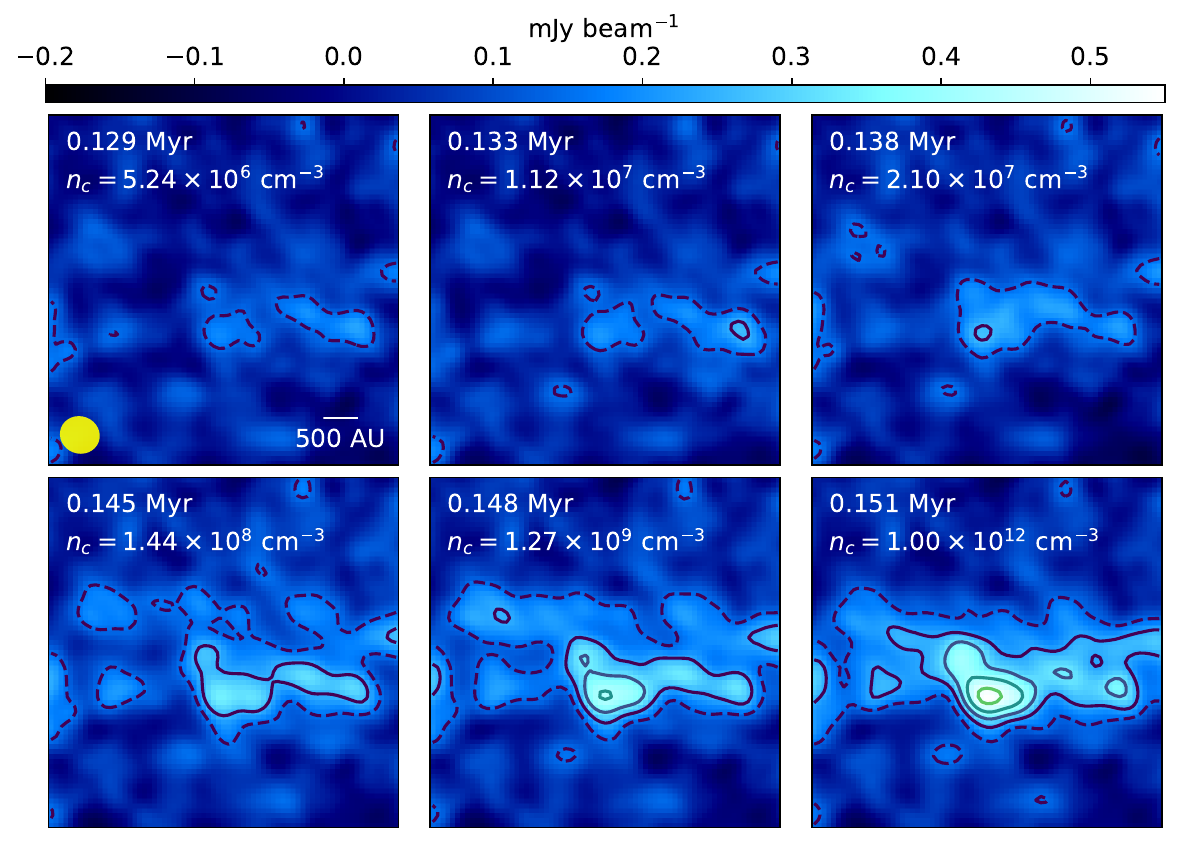}
    \caption{
    Synthetic ALMA 106~GHz observations of the $4M_{\odot}$ simulation, at six given timesteps, indicated in each panel along side the central density of the core. We adopt the same plotting convention as Figure \ref{fig:synthetic_alma}.
    \label{fig:synthetic_alma_4m}
    }
\end{figure*}

The starless core population in Orion B North, as catalogued by \citetalias{Nutter2007}, shows a typical mass of $1.4M_{\odot}$. Similar to \citet{Dunham2016}, we run synthetic observations on simulation snapshots of a $4M_{\odot}$ starless core and it's evolution. Figure \ref{fig:synthetic_alma_4m} shows that even in the case where the initial mass changes by a factor of 10, the detectability occurs at a similar central density threshold. \citet{Dunham2016} explore a similar set of results, and find that although the initial mass is vastly different between the two simulations, the amount of mass on compact scales, in this case approximately 1000~au, are similar. Additionally, due to the larger distance of 419~pc to the Orion B North cloud compared to Chamaeleon I (140~pc), there is overall less change within the primary beam over consecutive timesteps, leading to such a increase in detectability in the later timesteps shown.

\section{Figures}
\label{sec:additional-figures}

Although continuum images were made for all 73 individual fields observed, only a total of 19 individual fields yielded positive detections, of which many have been mosaicked together for better sensitivity. Below, we show all the remaining continuum detections found in our analysis.

\begin{figure*}
    \gridline{
        \fig{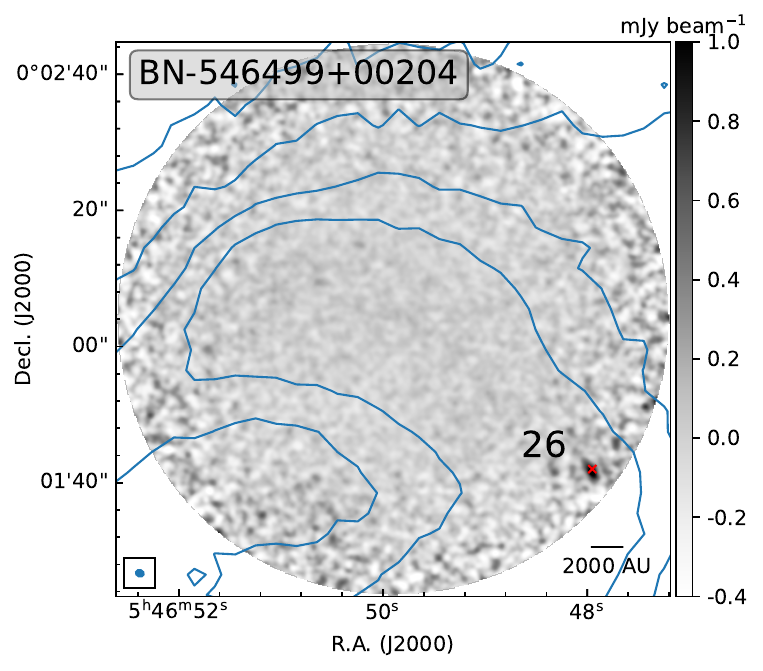}{0.5\textwidth}{}
        \fig{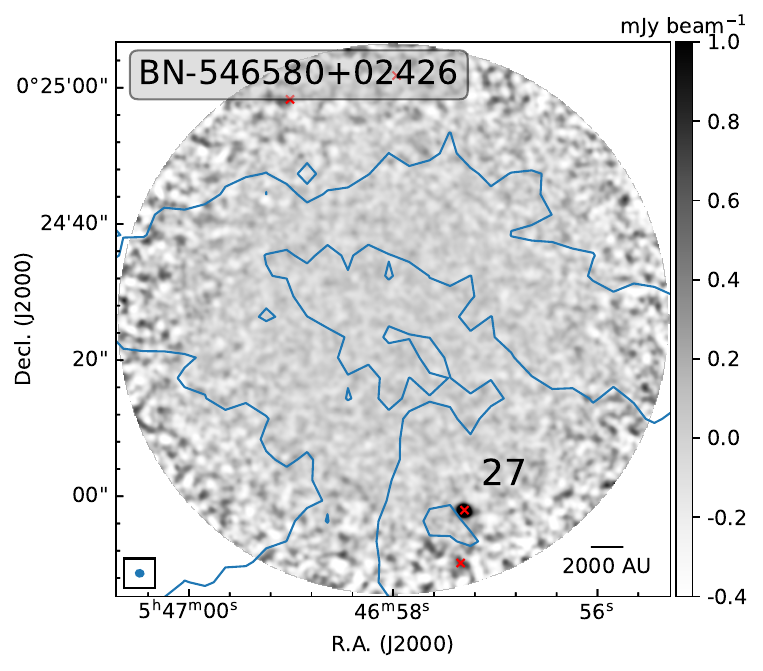}{0.5\textwidth}{}
        }
    \gridline{
        \fig{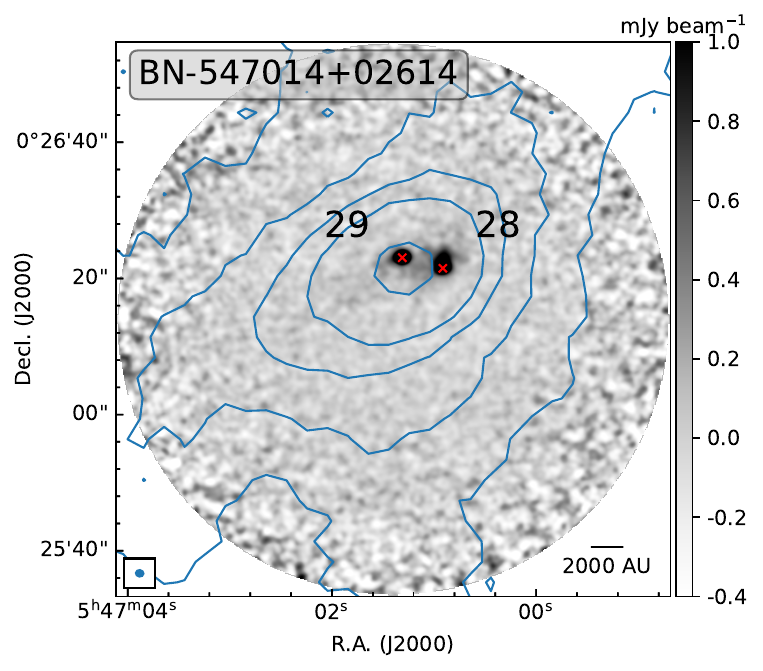}{0.5\textwidth}{}
        \fig{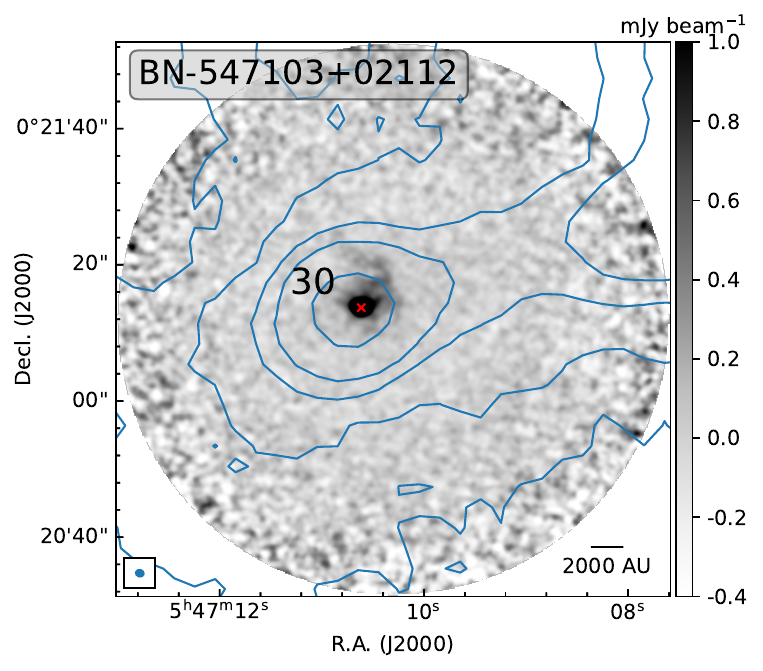}{0.5\textwidth}{}
    }
    \caption{
    ALMA single pointing fields that harbor detections. See Figure \ref{fig:Field0911} for plotting conventions. All detections in these individual pointings have an associated protostellar association.
    \label{fig:almadetections-4individuals}
    }
\end{figure*}

\begin{figure*}
    \gridline{
    \fig{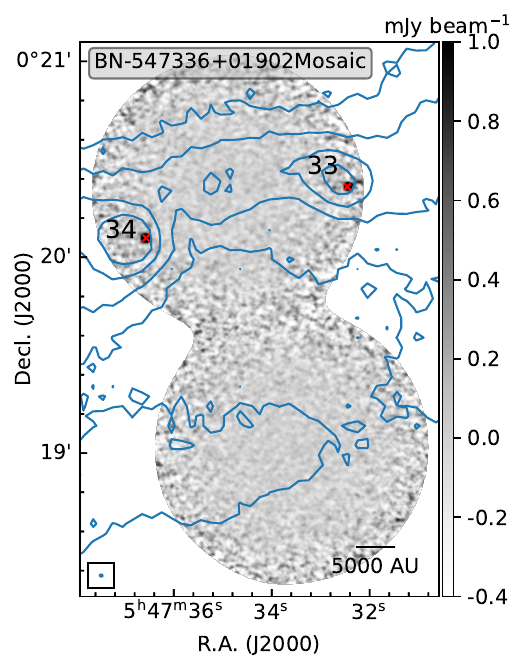}{0.5\textwidth}{}
    \fig{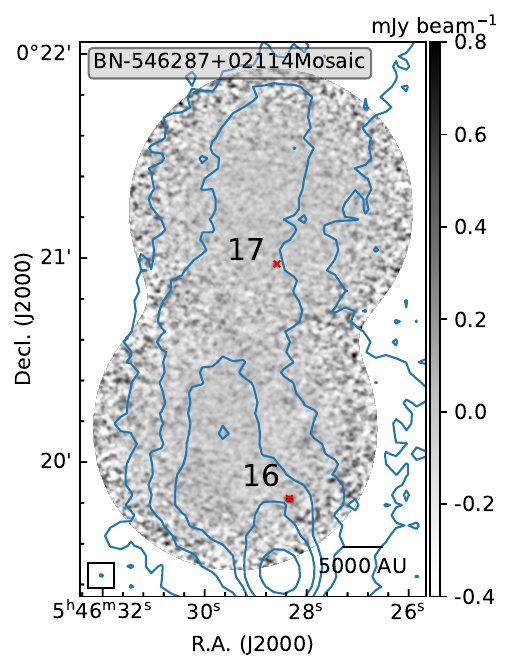}{0.5\textwidth}{}
    }
    \caption{
    ALMA mosaic fields that harbor detections. See Figure \ref{fig:Field0911} for plotting conventions. All detections in these mosaic pointings have an associated protostellar association.
    \label{fig:almadetections-1902-2114}
    }
\end{figure*}

\begin{figure*}
    \epsscale{0.75}
    \plotone{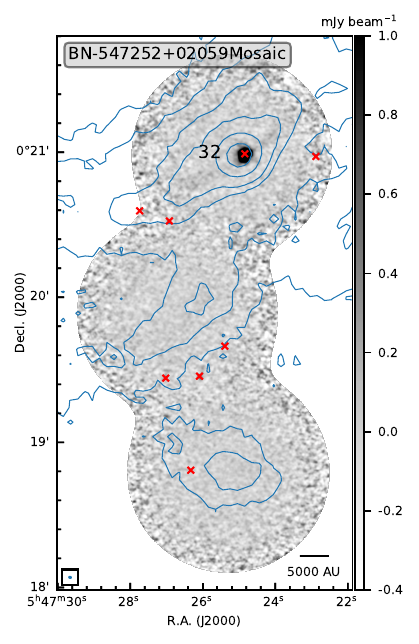}
    \caption{
    ALMA mosaic field that harbor detections. See Figure \ref{fig:Field0911} for plotting conventions. All detections in this mosaic pointing have an associated protostellar association.
    \label{fig:almadetections-2059}
    }
\end{figure*}

\begin{figure*}
    \gridline{
    \fig{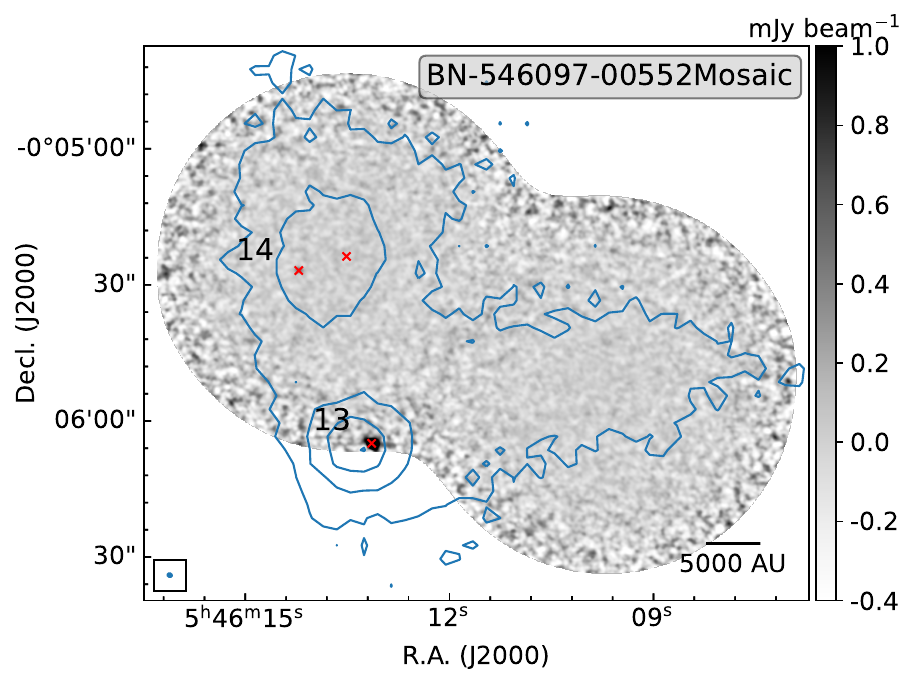}{0.58\textwidth}{}
    }
    \gridline{
    \fig{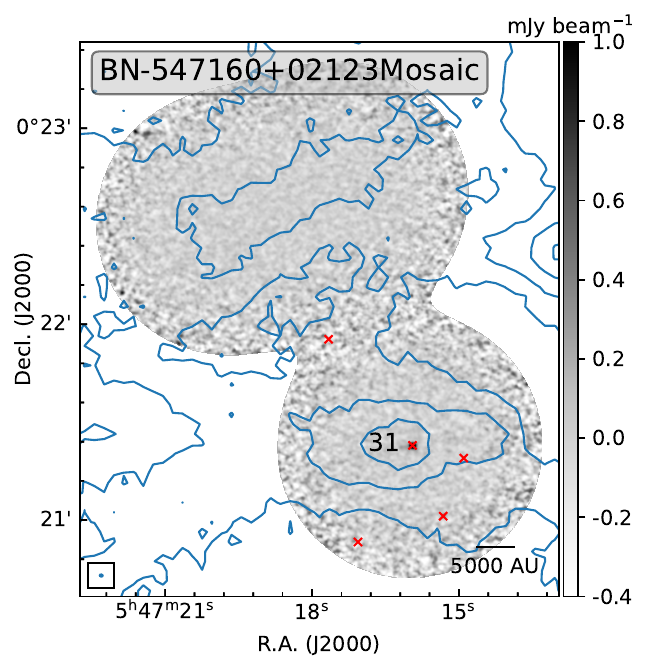}{0.58\textwidth}{}
    }
    \caption{
    ALMA mosaic fields that harbor detections. See Figure \ref{fig:Field0911} for plotting conventions. All detections in these mosaic pointings have an associated protostellar association.
    \label{fig:almadetections-0552-2123}
    }
\end{figure*}

\begin{figure*}
    \epsscale{0.8}
    \plotone{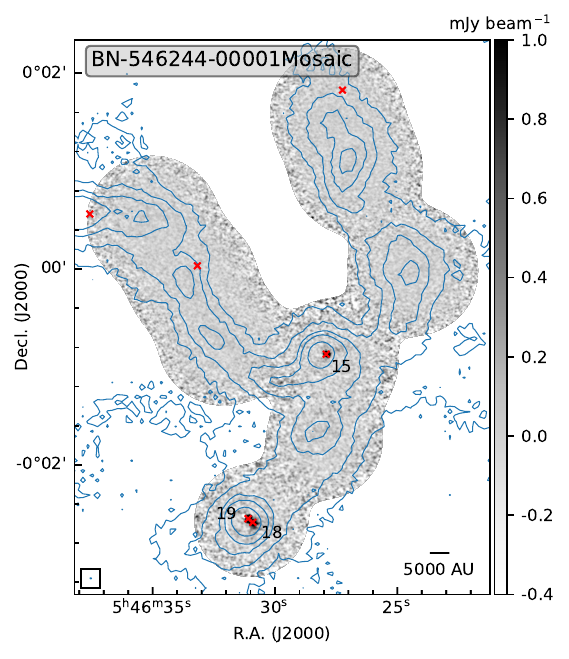}
    \caption{
    ALMA mosaic field BN-546244-00001 that harbor detections. See Figure \ref{fig:Field0911} for plotting conventions. All detections in this mosaic pointing have an associated protostellar association.
    \label{fig:almadetections_00001}
    }
\end{figure*}

\bibliography{main}{}
\bibliographystyle{aasjournal}

\end{document}